\documentclass[useAMS,usenatbib]{mnras}

\usepackage[normalem]{ulem}
\usepackage[T1]{fontenc}

\DeclareRobustCommand{\VAN}[3]{#2}
\let\VANthebibliography\thebibliography
\def\thebibliography{\DeclareRobustCommand{\VAN}[3]{##3}\VANthebibliography}

\usepackage{graphicx}	
\usepackage{amsmath}	
\usepackage{graphicx,colortbl}
\usepackage{txfonts}
\usepackage{pdflscape,lscape}
\usepackage{graphicx}
\usepackage{rotating}
\usepackage{newtxtext,newtxmath}
\usepackage{soul}

\newcommand{\degree}{$^{\circ}$}
\newcommand{\km}{${\rm km}\,{\rm s}^{-1}$}
\newcommand{\halpha}{H$_{\alpha}$}
\newcommand{\hi}{H\,{\sc i}}
\newcommand{\hii}{H\,{\sc ii}}
\newcommand{\angs}{${\textrm{\AA}}$}
\newcommand{\msolar}{${\rm M}_{\odot}$}
\newcommand{\msolaryr}{M$_{\odot}$\,yr$^{-1}$}

\title[Catalogue of dwarf galaxies]{Catalogue of nearby blue and near-solar gas metallicity SDSS dwarf galaxies}

\author[Yan Guo et al.]{
Yan Guo,$^{1,2}$ Chandreyee Sengupta,$^{1}$\thanks{E-mail: sengupta.chandreyee@gmail.com}
Tom. C. Scott,$^{3}$ Patricio Lagos$^{3}$ and Yu Luo$^{1,2}$
\\
$^{1}$ Purple Mountain Observatory,Chinese Academy of Sciences, 10 Yuanhua Road, Nanjing,  Jiangsu 210023, China  \\
$^{2}$ School of Astronomy and Space Sciences, University of Science and Technology of China, 96 Jinzhai Road, Hefei, Anhui 230026, China \\
$^{3}$ Institute of Astrophysics and Space Sciences (IA Porto), Rua das Estrelas, 4150-762, Porto, Portugal \\
}

\date{Accepted XXX. Received YYY; in original form ZZZ}

\pubyear{2015}

\begin{document}
\label{firstpage}
\pagerange{\pageref{firstpage}--\pageref{lastpage}}
\maketitle

\begin{abstract}
A less explored aspect of dwarf galaxies is their metallicity evolution. Generally, dwarfs have lower metallicities than Hubble sequence late--type galaxies but in reality, dwarfs span a wide range of metallicities with several open questions regarding  the formation and evolution of the lowest and the highest metallicity dwarfs.
We present a catalogue of  3459 blue, nearby, star--forming dwarf galaxies extracted from SDSS DR--16  including calculation of their metallicities  using  the mean of several calibrators. To compile our catalogue we applied redshift, absolute magnitude, stellar mass, optical diameter, and line flux signal–to–noise criteria.  This produced a catalogue  from the upper end of the dwarf galaxy stellar mass range. Our catalogued dwarfs have blue g -- i colours and H$\beta$ equivalent widths, indicative of having undergone a recent episode of star--formation, although their star--formation rates (SFR)  suggest only a moderate to low enhancement in star--formation, similar to  the SFRs in low--surface--brightness and evolved tidal dwarfs. While the catalogued dwarfs cover a range of metallicities,  their mean metallicity is $\sim$0.2 dex below solar metallicity,  indicating relatively chemically evolved galaxies. 
The vast majority of the catalogue, with clean photometry,  are relatively isolated dwarfs with only modest star--formation rates and a narrow range of g -- i colour, consistent with internally driven episodic mild bursts of star--formation. The presented catalogue's robust metallicity estimates for nearby SDSS dwarf galaxies will help target future studies to understand the physical processes driving the metallicity evolution of dwarfs.

\end{abstract}




\section{Introduction}
\label{intro}

Dwarf galaxies are overwhelmingly the most numerous type of galaxy in the universe, but due to their low luminosity, only the most luminous are detectable beyond the local Universe \citep{dale09,mccon2012}. Typically, dwarfs have much smaller optical diameters, 0.1 kpc to 10 kpc  \citep{deV91} and stellar masses,  10$^7$ \msolar\ to 10$^9$ \msolar\  \citep{lee06}, than the more massive galaxies in the Hubble sequence. 
Most dwarf galaxies also have a low oxygen abundance, \citep[12 + log(O/H) 7.0 to $\lesssim$ 8.4; e.g.][]{izotov99,kunth00,lee03}.  

Additionally, dwarf galaxies have been classified into several morphological subtypes, e.g., dwarf elliptical
galaxies (dE), dwarf spheroidal galaxies (dSph), dwarf irregular galaxies (dIrr),  dwarf spiral galaxies (dS), Magellanic type dwarfs (Sm, Im), blue compact dwarf (BCD) or HII \textcolor{blue}{(dwarf)} galaxies, ultra--compact dwarf galaxies (UCD), and TDGs, each with their characteristic properties \citep[e.g.,][]{sargentsearle1970,Pustilnik01,grebel01,grebel2003,grebel04, duc12,lisker13,janz17,arhn17}. 

There remain many unanswered questions about the origin and evolution of dwarf galaxies. For example, are the properties of the dwarf subtypes the result of different formation pathways or  do they reflect different stages of a dwarf galaxy's evolution?, e.g., 
\cite{Papaderos1996,vanZee1998,mayer01,sawala12,kim20}. While, for a few dwarf subgroups, e.g. TDGs, their origin and likely evolution are understood in outline at least \citep[e.g.,][]{duc97,braine,duc12,seng13,seng15,sengupta2017,scott18}, for most dwarfs  the physical processes driving their origin and relation to other dwarf types remain to be definitively established. Although, there are many studies which show correlations between dwarf properties (including kinematics, stellar populations and morphologies) and their environments \citep[e.g.,][]{lisker13,Sybilska18,toloba18}.  An interesting aspect of dwarf galaxy evolution is their metallicity evolution, in particular the relationship between the mass of stars in a galaxy and its metallicity \citep[e.g.,][]{tremonti04,berg12,jimmy15}.  Gas metallicity is the abundance of elements heavier than hydrogen and helium, so its value depends on the evolutionary state of the gas in the galaxies. A galaxy's gas--phase metallicity, Z, can be represented by 12 + log(O/H), where the solar metallicity
(Z$_{\odot}$) is 12 + log(O/H)$_{\odot}$ = 8.69 \citep{asplund2009}. 
The lower metallicity of dwarfs is generally attributable to retarded star formation  (SF) compared to more massive galaxies as a result of  the downsizing phenomenon.  There are however,  exceptions to the typical dwarf metallicity value, e.g., extremely metal--poor (XMP) dwarfs \citep[12 + log(O/H) $\lesssim$ 7.65,][]{Izotov2007,papaderos2008,lagos2014,lagos2016,lagos2018} and at the other extreme dwarfs with near solar oxygen abundances, e.g., TDGs \citep{duc99}.  It is still an open question whether the lowest metallicity dwarfs are survivors from the first epoch of galaxy formation or are young recently formed galaxies \citep[e.g.,][]{sargentsearle1970} or result from some other process such as metal--poor gas accretion \citep[e.g.,][]{lagos2016,lagos2018}.  To date, few studies of high metallicity dwarfs are available, and as a result, their formation scenarios are yet to be explored systematically. TDGs are known to be high--metallicity dwarfs,  but not all high metallicity dwarfs have been unambiguously found to have a tidal origin \citep{sweet2014}.

 While studies of the metallicity evolution in dwarf galaxies exist in the literature, they are either  detailed, resolved studies of small samples
\citep[e.g.][]{lee06,lagos2007,berg12,jimmy15}, or larger samples with additional selection criteria like a detected \hi\ counterpart or limited range in metallicity \citep[e.g.][]{jimmy15,James2017} or
higher redshift dominated samples \citep{tremonti04,
peebles08,li23}. An optically selected statistically large sample of low redshift dwarf galaxies with robust metallicity estimates has not previously been available in the literature. This paper aims to
fill that gap and presents a dwarf galaxy catalogue with metallicity estimates. The catalogue will facilitate the selection of subsamples for future multi--wavelength follow--up or statistical studies to identify the physical processes driving the metallicity
evolution in specific groups of dwarfs, especially for high metallicity dwarfs.  

We initially compiled a catalogue of undifferentiated dwarf galaxies from the Sloan Digital Sky Survey (SDSS) DR--16 \citep{york00}.   To quantify dwarf galaxy properties requires accurate redshifts. Therefore, we restricted our initial selection to galaxies with SDSS spectra, which allowed their redshifts to be determined accurately.  Galaxies from this initial selection had the potential for other studies based on their spectra, e.g., determination of their metallicity, and star formation history.  But these types of study require galaxies which present strong emission lines. The final dwarf catalogue was restricted to dwarfs with sufficiently strong emission lines in their SDSS spectra to robustly determine their metallicities.  Because of the selection criteria,  the catalogue is not representative of dwarfs in general, but instead, it consists of the more massive star--forming, relatively high metallicity dwarfs in SDSS.  

The layout of the paper is as follows. In Section \ref{methods} we discuss the methodology used to compile the catalogue. Section \ref{sec:result} presents the results obtained (e.g.,  metallicity estimates, star--formation rate (SFR) and several other properties of the catalogue). Section \ref{discuss} is a discussion on the scientific implications of the results and finally Section \ref{sec:conclude} summarises the work carried out and the main results of the paper. 

\section{Methodologies}
\label{methods}
\subsection{The Sample}
\label{sample}
   
\begin{figure*}
\includegraphics[scale=.45]{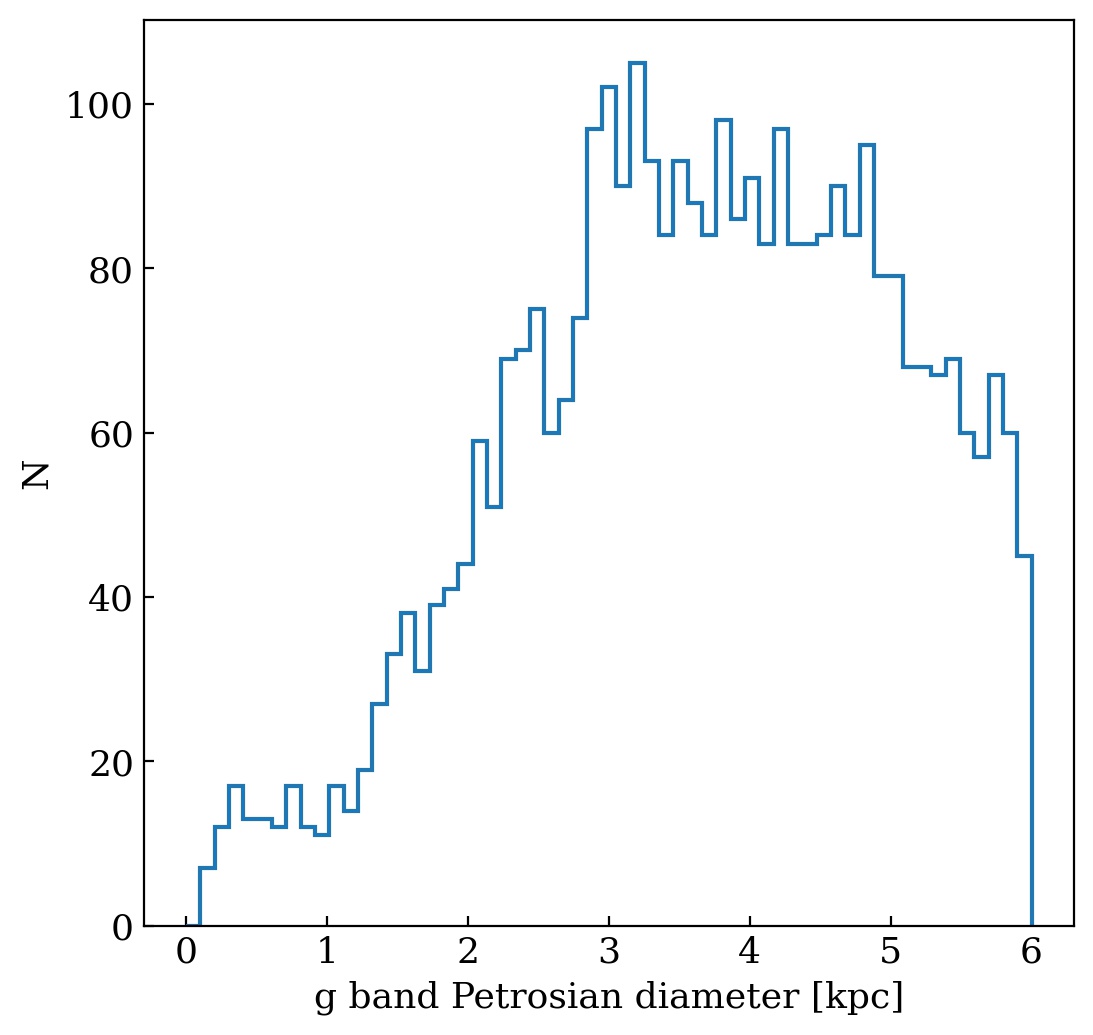}
\includegraphics[scale=.45]{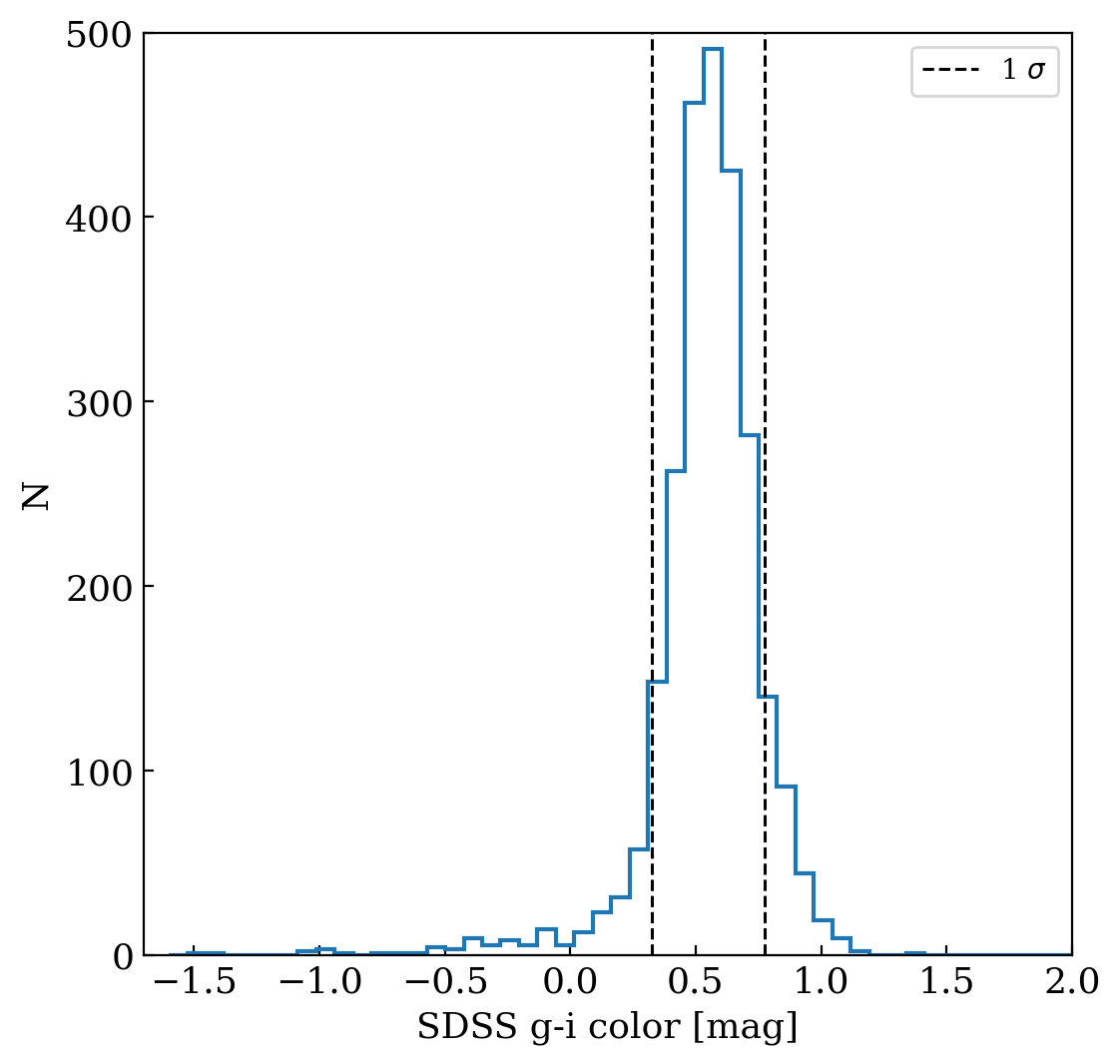}
\includegraphics[scale=.45]{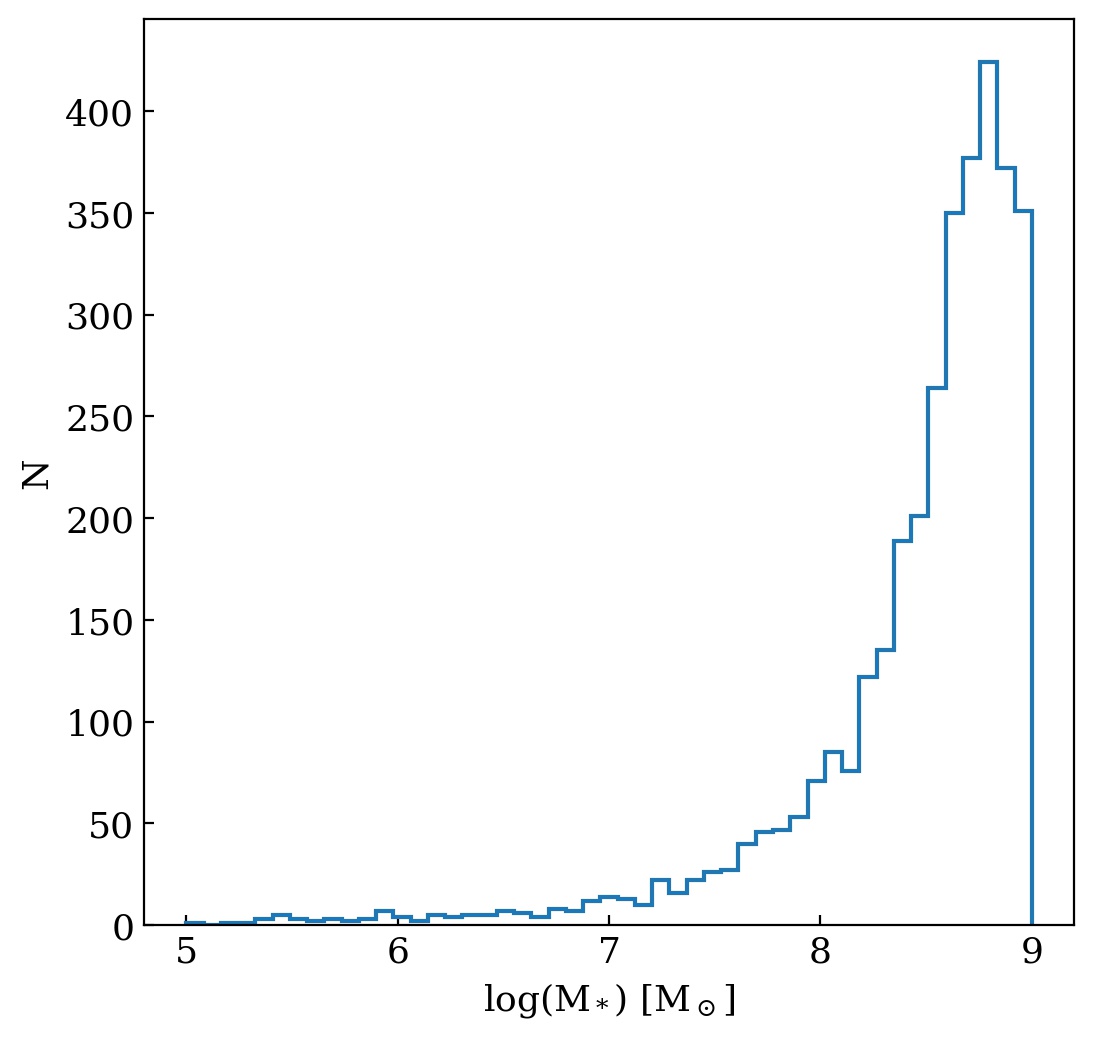}
\includegraphics[scale=.45]{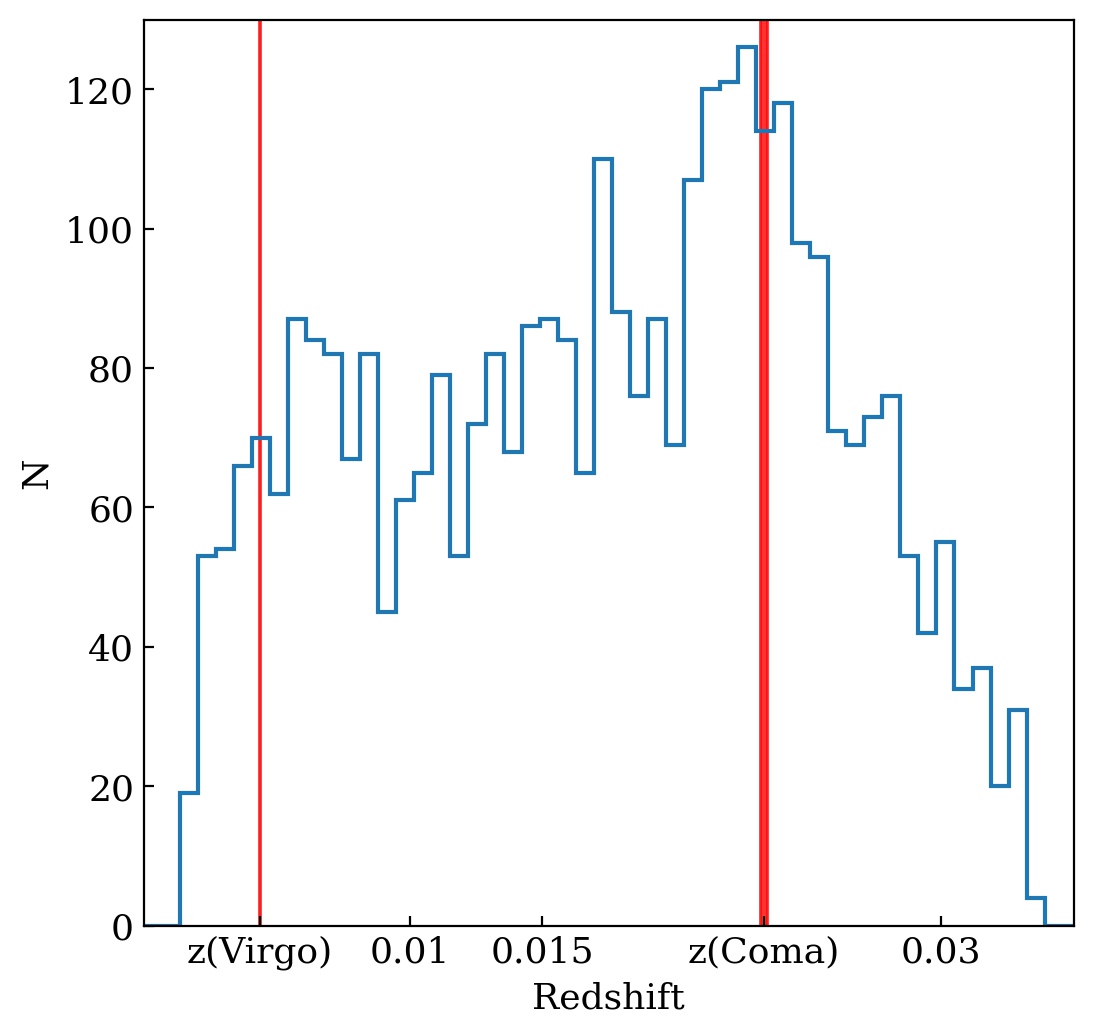}
\caption{\textcolor{black}{Histograms of selected properties of the catalogued dwarf galaxies: \emph{Top left:} D$_{pet}$ in kpc, \emph{Top right:} SDSS g -- i colour, \emph{Bottom left:} log(M$_*$/\msolar), \emph{Bottom right:} redshift, with the vertical red lines indicating the redshift and redshift uncertainties of the   Virgo (z = 0.00436$\pm$0.00002) and Coma (z = 0.02333$\pm$0.00013) galaxy clusters.}}
\label{fig:result}
\end{figure*}

We started by selecting all DR--16  SDSS spectra within \textcolor{black}{the} radial velocity range of 500 \km\  $\leq$ v$\leq$ 10000 \km, with a 'GALAXY' classification in 'Class', 'Type' and/or 'SourceType'. This selection eliminated stars and quasars from our source list. The velocity range was chosen to avoid the difficulty of establishing distances to sources at velocities below 500 \km\ and   the limited number of dwarfs with spectra in SDSS at velocities $>$10000 \km. This \textcolor{black}{upper limit} is confirmed in Figure \ref{fig:result}, bottom right panel, where we see that the number of dwarfs in the catalogue falls rapidly toward the catalogue's redshift upper limit.

To eliminate massive galaxies, all galaxies with an absolute g -- band ($\sim$ 4770 \angs)  magnitude, M$_{g}$  $\le$ -18 mag were excluded from our sample. Based on  this selection of galaxies with absolute magnitudes  $\ge$ -18 criteria, we  obtained a preliminary catalogue containing  18768 dwarf galaxy candidates.
The next criterion applied was galaxy stellar mass. The stellar masses of the  18768 galaxies were estimated using the relationship between \textcolor{black}{the candidates' photometric SDSS DR--16 model magnitudes for selected colour band filters and  the stellar M/L ratio from \cite{bell03}. The SDSS Pestrosian g -- band luminosity ($L_g$) and $g - r$ colour of the sources was used to calculate the }  stellar mass (M$_*$) using Equation \ref{eq:1}:  

\begin{equation}\label{eq:1}
    log(M_*/L_g) = -0.499+1.519\left(g - r\right)
\end{equation}

\textcolor{black}{For our dwarf catalogue, we chose only galaxies with stellar masses less than 10$^{9}$ \msolar, reducing the previous candidate} list of  18768 galaxies to 13211. The above selection steps ensured that the catalogue contained only low stellar mass galaxies, but could not ensure that they were necessarily dwarf galaxies. To exclude galaxies with large optical diameters, such as low surface brightness galaxies,  we set an upper D$_{25}$ limit of 10 kpc, approximately the diameter of the Large Magellanic Cloud  \citep{deV91}. We calculated the relation between the SDSS Petrosian (R$_{pet}$ $\times$ 2) diameter (D$_{pet}$) and the  D$_{25}$ isophotal diameter for a sample of the catalogue's galaxies and found D$_{pet}$ = 0.67D$_{25}$ (D$_{25}$ $\sim$ 1.5D$_{pet}$). Thus we used a g -- band D$_{pet}$ cut off of 6 kpc, which is approximately equivalent to a D$_{25}$ of $\le$ 10 kpc.  The sizes of the galaxies in kpc were obtained  using the kpc/arcsec scale from the online Ned Wright Cosmology Calculator assuming H$_0$ = 67.8 \km Mpc$^{-1}$, $\Omega_M$ = 0.308 and $\Omega_{vac}$ = 0.692 \citep{wright2006}. The D$_{pet}$ values in arcseconds were derived from SDSS.  The catalogue of dwarf galaxies after applying the optical size cut--off contained  5683 sources. A further 55 candidates were removed from the catalogue as they displayed a clear spiral structure, inconsistent with them being dwarf galaxies. Amongst the remaining  5628 galaxies, 26  had multiple spectra for the same galaxy but from different projected positions. In these cases, we only retained the spectrum closest to the galaxy centre and thus our initial dwarf galaxy catalogue then contained  5602 candidates.

Having established  an initial dwarf galaxy selection we further  refined the catalogue by selecting  only those dwarfs displaying strong emission lines in their SDSS spectrum  which facilitated  the study of their metallicity and star formation  histories. 
To do this, first, we removed the stellar component following the procedure explained in Section \ref{sec:Spectral_syn}. Then we  chose a subset of dwarfs based on the signal--to--noise ratio (SNR)  of the spectral lines needed to determine their metallicities. To increase the robustness of the metallicity and other estimates (e.g., electron density), we only chose candidates with spectra where the emission lines used to determine these quantities (H$\beta$ 4861\angs, [O III] 5007\angs, H$\alpha$ 6563\angs, [N II] 6584\angs, [S II]6717\angs,[S II] 6731\angs)  all had  a SNR $\ge$ 2.5. The SNR criterion was not applied to the weak  [O III] 4959\angs\ and 4363\angs\ lines, which were detected in only some galaxies. Applying this  emission line SNR  criterion reduced the number of galaxies in our final dwarf catalogue to 3459.   Using the final  3459 galaxies, the four panels of Figure \ref{fig:result} show the number distribution of the catalogue  with respect to Petrosian diameter, g -- i colour, log(M$_*$/\msolar)   and redshift.

\subsection{Spectral fitting}
\label{sec:Spectral_syn}

Spectral synthesis is an important method to derive galaxy properties from their spectra. \textcolor{black}{To fit the stellar component of our  catalogued dwarfs we used 
the spectral synthesis code Fitting Analysis using Differential evolution Optimization
\cite[FADO,][]{gomes17}}.

One important issue to be considered during the spectral fitting is the effect of nebular continuum on the stellar SED. 
For a galaxy with a low  SFR, this does not significantly impact the analysis of the spectrum. However,  in star--forming galaxies the nebular continuum can contribute significantly in determining the SED and stellar properties (ages, M$_*$, etc.) of galaxies \citep{cardoso22}. Removing the stellar component from each spectrum then allowed us to quantify the  spectrum's the emission line fluxes    without the nebular continuum contribution.
For our catalogue, this issue is relevant because  our catalogue includes some strongly star--forming low mass galaxies, i.e., BCDs.

\begin{figure*}
\includegraphics[scale=.60]{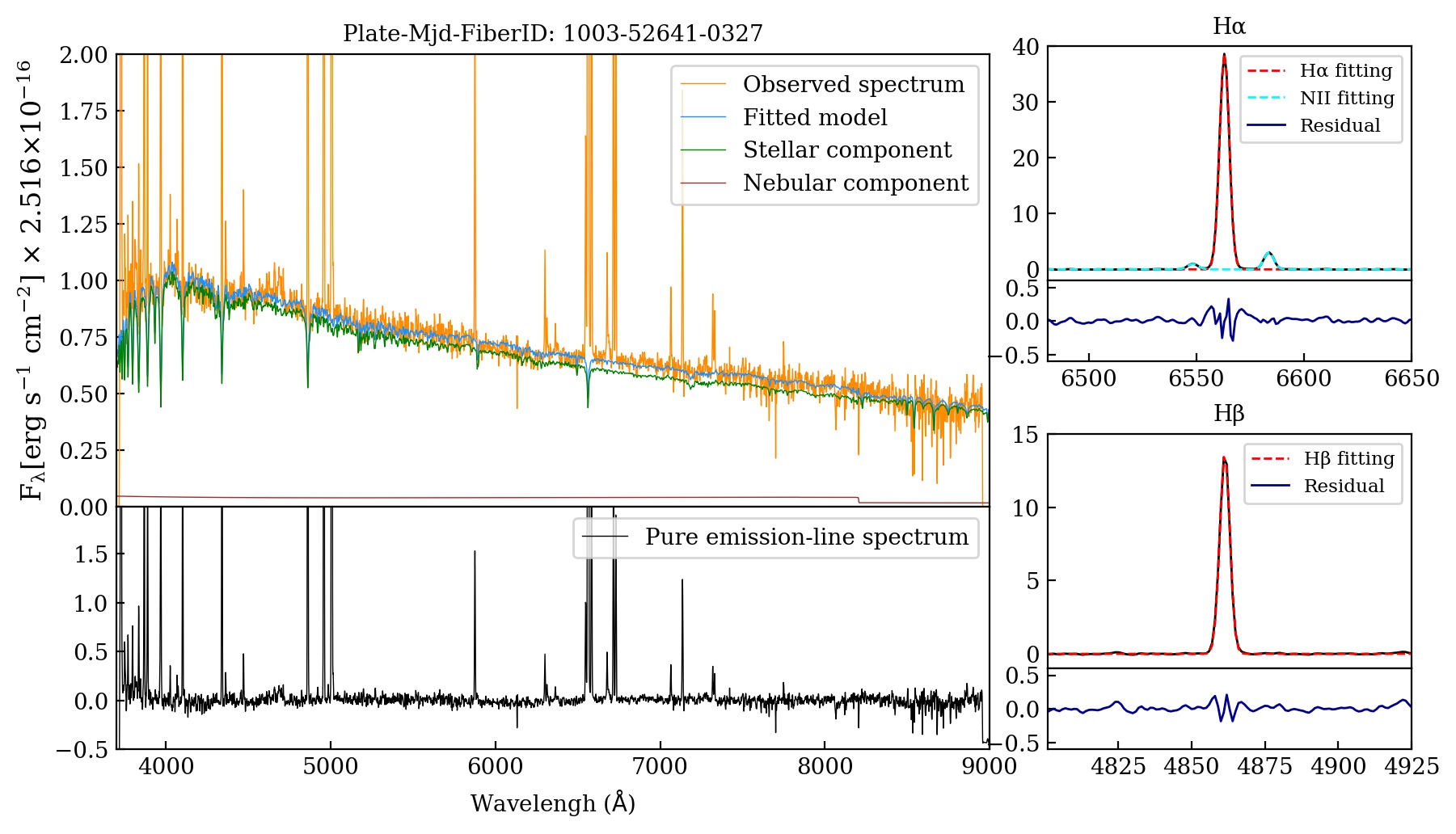}
    \caption{Example of FADO continuum (stellar and nebular) fits to an SDSS dwarf galaxy spectrum   with the fits colour coded per the legend (top left) and the derived pure emission line spectrum (bottom left).  The right upper and lower right panels, respectively, show the fitted H$\alpha$ and H$\beta$ lines, with the residuals shown at the bottom of each panel.  The SDSS spectrum is from Plate = 1003, MJD = 52641  FiberID = 0327.}
    \label{fig:fado}  
\end{figure*}

FADO reproduces the observed nebular characteristics of a galaxy and ensures consistency between the best--fitting stellar model and the observed nebular emission characteristics to a  greater extent than other codes. For these reasons, we chose to apply FADO to measure the line fluxes of our dwarf catalogue candidates. Figure \ref{fig:fado} shows an example of FADO's fits to the SDSS spectrum of a dwarf galaxy in our catalogue. Before running FADO, the SDSS  spectra were corrected for Galactic  A$_V$ foreground extinction using the extinction map from \cite{1998ApJ...500..525S} in the NASA/IPAC Extragalactic Database (NED) and the \cite{Cardelli1989} extinction law with R = 3.1. We ran FADO V1.B using the simple stellar population (SSP) templates from the \cite{bruzualCharlot2003} libraries with metallicities of Z = 0.004, 0.008, 0.02 and 0.05 and stellar ages between 1 Myr and 15 Gyr, assuming a Chabrier initial mass function \cite[IMF,][]{Chabrier2003}. The fit was performed in the \textcolor{black}{3800 -- 9200}\angs\  spectral range using the \cite{Cardelli1989} extinction law. Here, we assumed Z$_{\odot}$ = 0.02.

 Finally, the chosen wavelength range \citep[see][]{cardoso22} includes the noisier parts of the SDSS spectra. To see if the inclusion of these noisier parts of the spectra  affected the robustness of the implicit stellar populations and,  consequently, the determination of the galaxies' nebular properties we ran FADO limiting the spectral range to 4300 -- 7000\angs. Using this less noisy spectral range, we found that the mean stellar ages weighted by light and mass are 0.944 Gyr and 2.393 Gyr, respectively. While using the full wavelength range the mean stellar ages weighted by light and mass are 0.718 Gyr and 1.802 Gyr, respectively.  We also found that the mean nebular properties differed by, 2\%, 0\% and 4\% for the H$\alpha$ fluxes, log [NII]/H$\alpha$ and log [OIII]/H$\beta$, respectively. This implies  a difference of 0.01 dex for the mean 12 + log(O/H). In section \ref{result:metal} we   describe the methods used in this work to estimate the oxygen abundance. We conclude our chosen SDSS spectral range does not materially affect the results or interpretations of this study.

\subsection{Emission Line Flux}

\label{em_line_flux}
\begin{figure}
\includegraphics[width=\columnwidth]{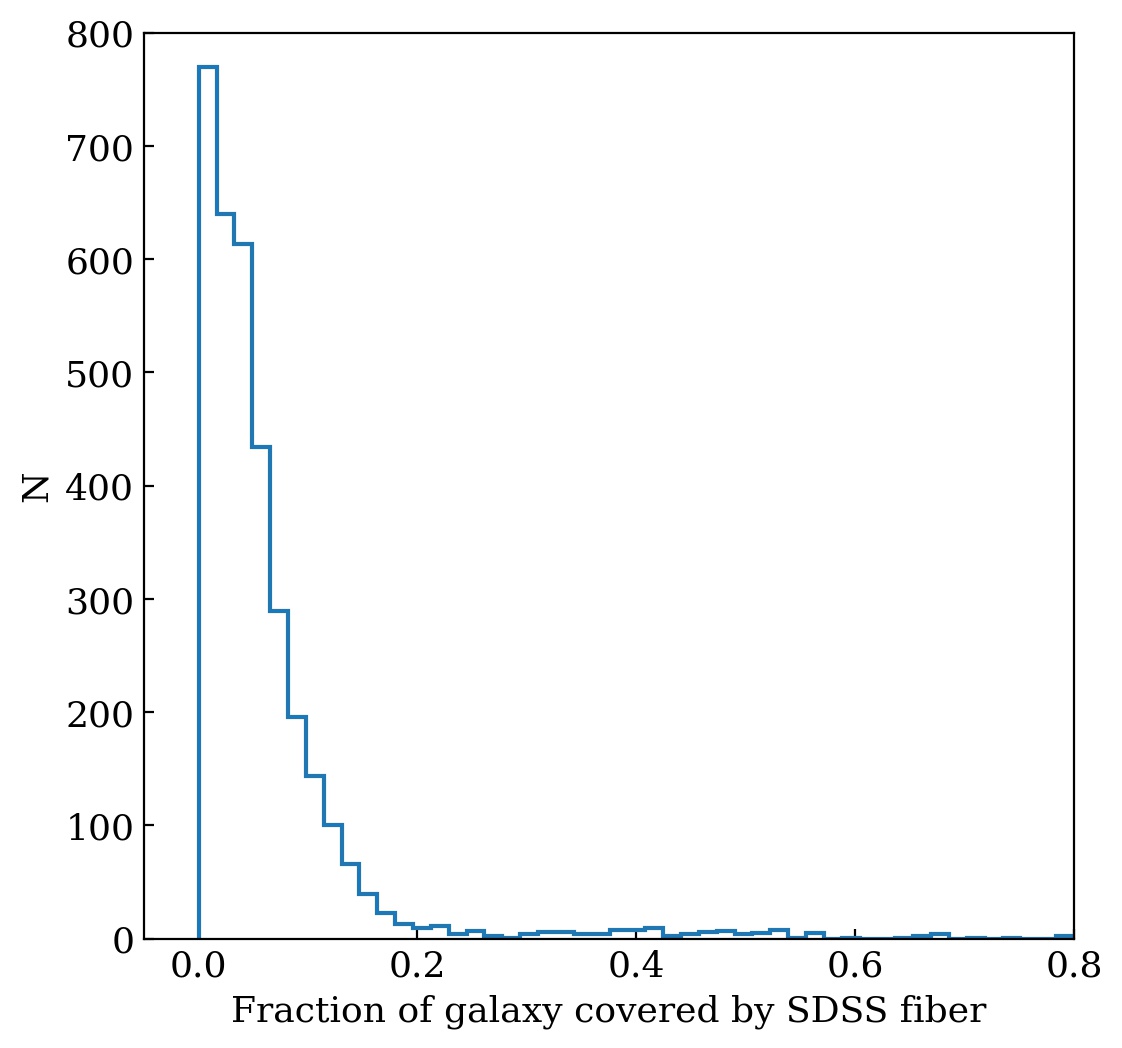}
\caption{Histogram showing the distribution of the fraction of the catalogued dwarfs' surface area covered by the SDSS spectral fibre (D$_{fiber}$ = 3 arcsec). Each galaxy's surface area was estimated from its SDSS g -- band D$_{25}$.}
\label{fig:1a}
 \end{figure}

Emission line fluxes were derived from analysis of the candidates' SDSS spectra. The SDSS spectra were derived from fibers with a three arcsec diameter within the wavelength range 3800\AA\ to 9200\AA. The fraction of a galaxy's optical emission  detected within this wavelength range depends on the  dwarf's angular diameter and  redshift. 
The catalogued dwarfs have a mean SDSS fibre coverage fraction of the dwarfs'  deprojected optical disks of approximately $\sim$0.07. This fraction  is based  on a surface area estimated from  the g -- band D$_{pet}$ $\times$ 1.5 which is approximately equal to D$_{25}$ (see Section \ref{intro}). Our calculation  assumes the galaxies are face--on and thus   provides a lower limit for the fractional fibre coverage. Figure \ref{fig:1a} shows the distribution of the fibre coverage fraction for the catalogued dwarfs'  surface area based on their D$_{25}$, extrapolated from  their g -- band D$_{pet}$. As  a cross--check, the fibre coverage fraction was re--calculated  using galaxy diameters from NED, which  also gave a mean coverage fraction of $\sim$ 0.07.  Using either method, we see the spectra are typically based on the emission from less than 10\% of each dwarf's surface area. However, the relatively small physical diameter of dwarfs and the short time scale for  metallicity homogenisation across  dwarfs \citep[][and references therein]{croxall09,lagos2009,lagos2012,lagos2013} means we have no reason to consider the metallicities derived from the SDSS spectra are unrepresentative of each dwarf's whole surface area as would be the case for larger spiral galaxies with distinct bulge and disk components.
We applied the FADO  code to the SDSS spectra of all potential members of our catalogue,  as detailed in Section \ref{sample}, to get pure and precise observed emission line fluxes. We then corrected for intrinsic extinction to derive  intrinsic emission-line fluxes from integrated observed flux F($\lambda$) using  \textcolor{black}{Equation \ref{eq:2}:}

\begin{equation}\label{eq:2}
    \frac{I(\lambda)}{I(H\beta)} = \frac{F(\lambda)}{F(H\beta)} \times 10^{c({H\beta})f(\lambda)}
\end{equation}

where f($\lambda$) is the reddening function taken from \cite{Cardelli1989} assuming R$_v$ = 3.1. The logarithmic reddening parameter c(H $\beta$) was calculated assuming case B  \citep{2006agna.book.....O} for the Balmer decrement ratio, H$\alpha$/H$\beta$ = 2.86 at 10000 K. 

As our catalogue is a subset of all galaxies with SDSS DR--16 spectra, it is useful to compare selected properties of our  catalogued galaxies with the SDSS value--added catalogue produced by MPA--JHU\footnote{\url{https://wwwmpa.mpa-garching.mpg.de/SDSS/DR7/}}. In Appendix \ref{comparison}, Figure \ref{fig:comparison_flux_ratios_fado_mpa} (left)  we show the comparison of H$\alpha$ and H$\beta$ fluxes for our dwarf galaxies. The x--axes show values derived from our methods and the same properties from MPA--JHU on the y--axis.
The emission line fluxes shown in Figure \ref{fig:comparison_flux_ratios_fado_mpa}, differ at 1$\sigma$ level from the MPA--JHU catalogue values. Similar results  to ours were obtained by \cite{Miranda2023}. This difference is a known issue, which arises from different flux measurement methods.  

\subsection{BPT diagnostic diagram}
\label{subsec:bpt}

\begin{figure}
\includegraphics[width=\columnwidth]{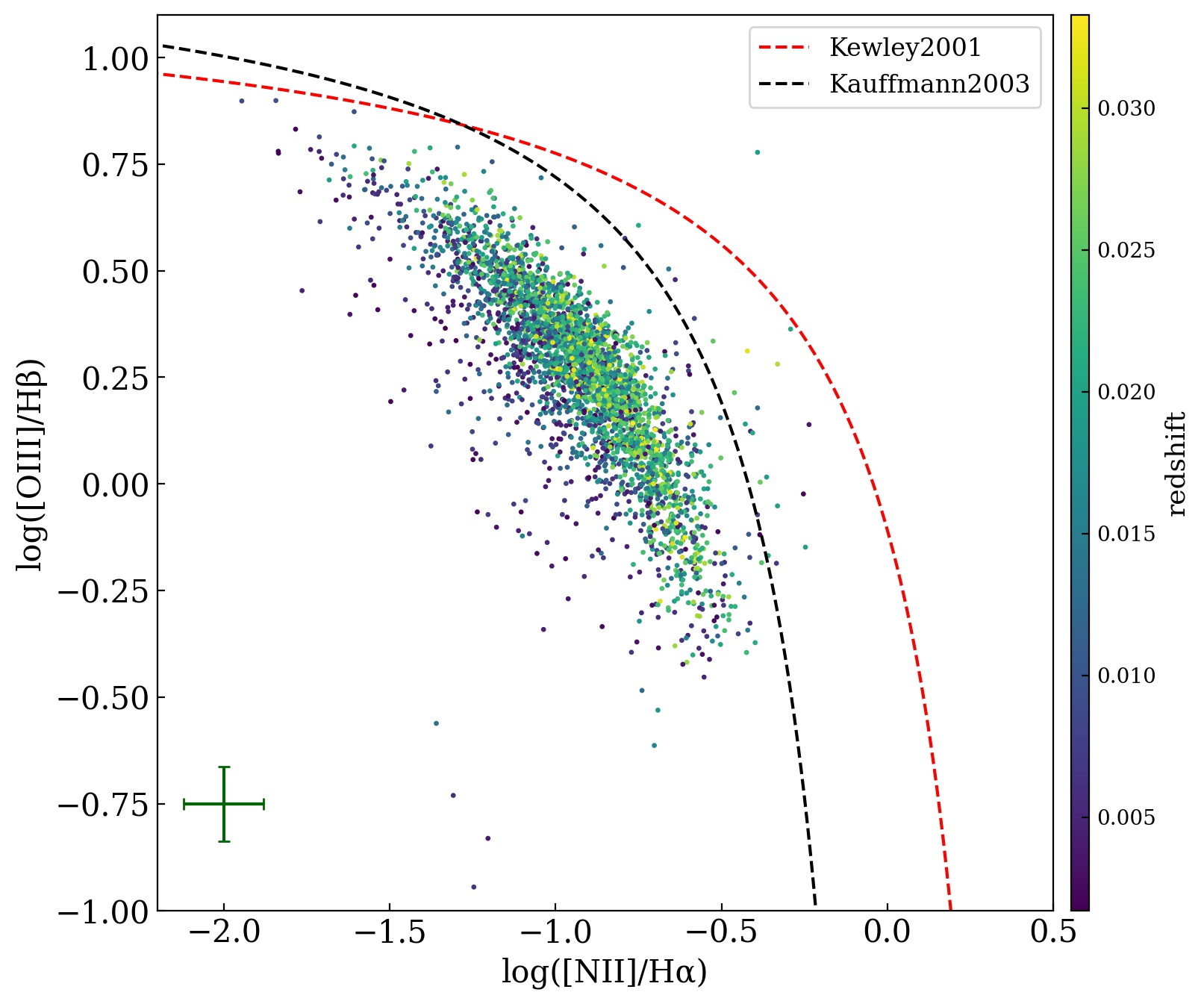}
\includegraphics[width=\columnwidth]{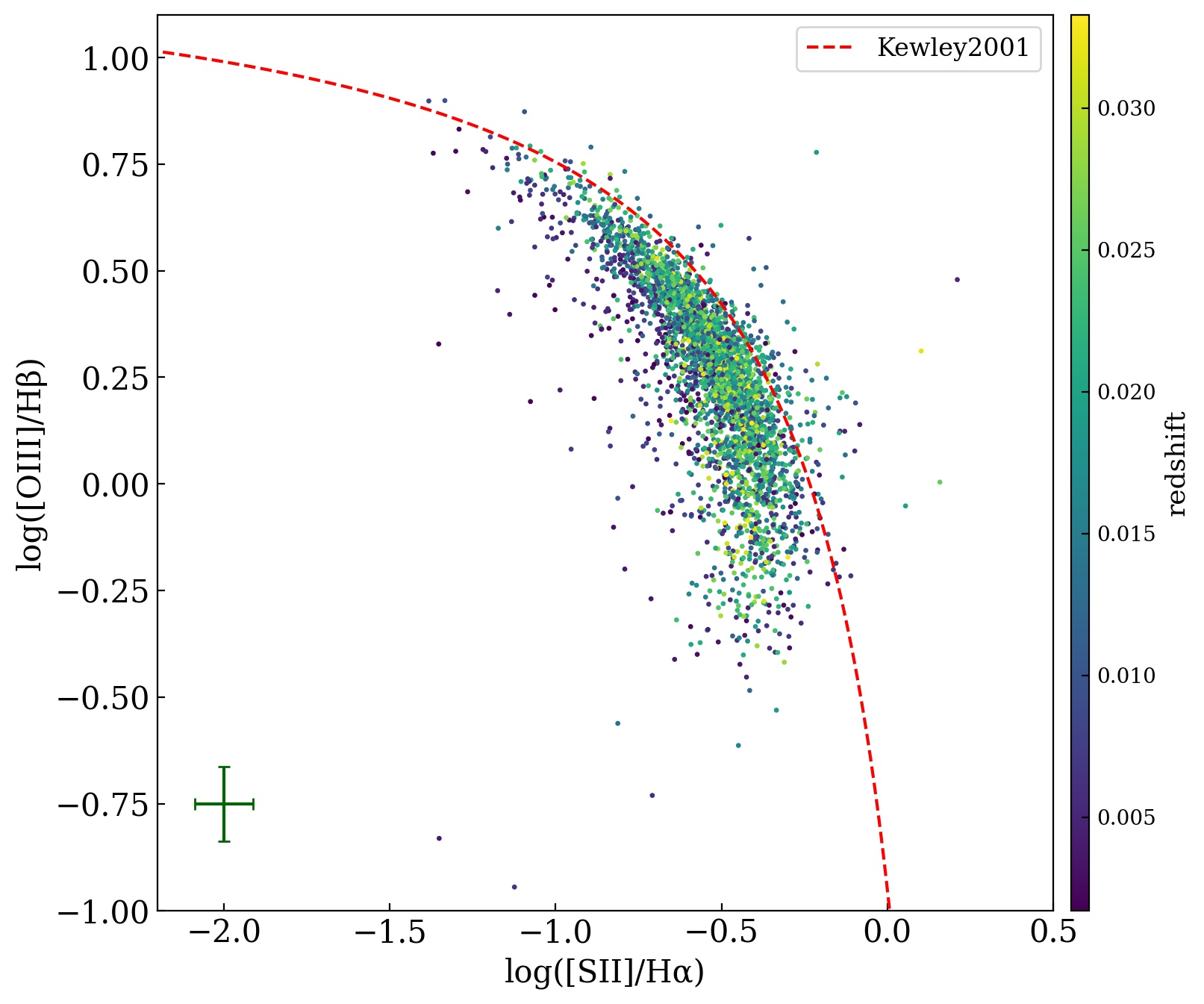}
\caption{BPT log([N II] $\lambda$6584/H$\alpha$) versus log([O III] $\lambda$5007/H$\beta$) (upper panel) and log([S II] $\lambda\lambda$6717,6731/H$\alpha$) versus log([O III] $\lambda$5007/H$\beta$) (lower panel)  diagnostic diagrams for the 3459 galaxies in the dwarf catalogue. 
We include, in the diagrams, the Kewley et al. (2001; red dashed line) and Kauffmann et al. (2003; black dashed line) model boundary lines which divide regions dominated by photoionisation, composite, and AGN/LINERs.}

\label{fig:bpt}
\end{figure}

Dwarf galaxies in our final catalogue consist of unclassified dwarfs and their spectra could potentially be dominated by an energy source other than stars, e.g.,  active galactic nuclei (AGN) or shocks.  In Figure \ref{fig:bpt} we use  Baldwin, Phillips, Terlevich (BPT) diagnostic diagrams \citep{1981PASP...93....5B}: log([N II] $\lambda$6584/H$\alpha$) versus log([O III] $\lambda$5007/H$\beta$), upper panel, and log([S II] $\lambda\lambda$6717,6731/H$\alpha$) versus log([O III] $\lambda$5007/H$\beta$),  lower panel, to determine the catalogued dwarfs' ionisation sources. The demarcation dashed curves in the figures are from the model by \cite{2001ApJ...556..121K} (red line) and \citep{Kaufmann03} (black line). The lower left areas of both panels are \hii\ regions dominated solely by photoionisation from hot young stars. The area at the top and right of the dashed red curves is inhabited by galaxies who's dominant ionisation mechanism is AGN and/or low--ionization nuclear emission--line region (LINER) emission. On the other hand the area between the two curves is the composite region, where both AGN/LINERs and SF contribute significantly to the ionisation. In the upper panel of Figure \ref{fig:bpt} we see that most galaxies fall in the locus predicted by models of photoionisation by young stars. However, there are a few exceptions i.e., the 23 galaxies in the composite region of the NII diagram and the one galaxy falling in the pure AGN region. Although, we did not detect asymmetric line proﬁles at the base of H$\alpha$ lines for these 23 dwarfs, future observations may reveal the presence of AGN in this small sub--sample of our catalogue. Similarly, in the SII BPT diagram, Figure \ref{fig:bpt} lower panel, 181 galaxies fall in the AGN dominated region.  The fraction of AGN dominated galaxies in both panels of the plot is very small.


\section{Results}
\label{sec:result}

\subsection{Metallicity Estimates}
\label{result:metal}

The BPT diagrams, in Section \ref{subsec:bpt},  show that the majority of the galaxies in our catalogue are local star--forming dwarf galaxies. This demonstrates our sample is  consistent with the samples used for calculating metallicity of local \hii\  galaxies \citep[][]{terlevich1991} in \cite{denicolo02,PettiniPagel2004,perez09}, where the 12 + log(O/H) abundance is characterized using \hii--based calibrators.
The direct T$_e$ method for determining a galaxy's metallicity is generally considered the most accurate and uses [O III] $\lambda$4363 and $\lambda\lambda$4959, 5007 lines and the [O II] $\lambda$3727 doublet to estimate oxygen abundance. However, the weak [O III] $\lambda$4363 auroral emission line is difficult to detect. Moreover,  in the redshift range of our candidates, the [O II] $\lambda$3727 spectral line appears within a noisy region  very close to the edge of the SDSS spectrometer wavelength range or is redshifted beyond the spectrometer's wavelength range. As a result we detect the lines needed to apply T$_e$ method in only about half of the galaxies in our catalogue. Because of these constraints the errors in applying the T$_e$ method can be large. 
Therefore, we use the alternative strong line calibrators to estimate the metallicities of the dwarf galaxies more consistently and accurately. We compute oxygen abundances using several of these diagnostic methods \textcolor{black}{(see the list of calibrators used in Table \ref{tab:met})} and the following indexes:

\begin{equation}
    N2 = \log(\frac{H\alpha}{[N II]\lambda6584}),
\end{equation}

and

\begin{equation}
    O3N2 = \log(\frac{[OIII]\lambda5007}{H\beta} \times \frac{H\alpha}{[NII]\lambda6584}),
\end{equation}

 We  summarise below the calibrators used in this study:
\begin{itemize}
\item The calibrations from \cite{PettiniPagel2004}, \cite{denicolo02}  and \cite{perez09}  come from  nearby \hii\ galaxies, using the T$_e$ method, and their sub--samples of dwarf irregular galaxies using mainly N2 and O3N2 indexes.
Some additional diagnostics:
\item  \cite{marino13} also used T$_e$--based measurements but for \hii\ regions from the CALIFA sample and 603 \hii\ regions from the literature.
\item \cite{bian18} use the spectra of a sample of local MPA--JHU SDSS analogues of high--redshift galaxies.

\item \cite{Naga06} also  used data from the SDSS but with known metallicity available from the literature, and gives an analytical description for the metallicity dependence of the many diagnostics and line ratios within the metallicity range 7.0 $\leq$ 12 + log(O/H) $\leq$ 9.2. 
\item \cite{Curti20} determined the metallicity via the $T_e$  method in SDSS star--forming galaxies with z > 0.027, the diagnostics proposed in that work were calibrated against metallicity in the range 7.6 $\leq$ 12 + log(O/H) $\leq$ 8.9. 
\end{itemize}

\begin{figure*}

   \includegraphics[scale=.41]{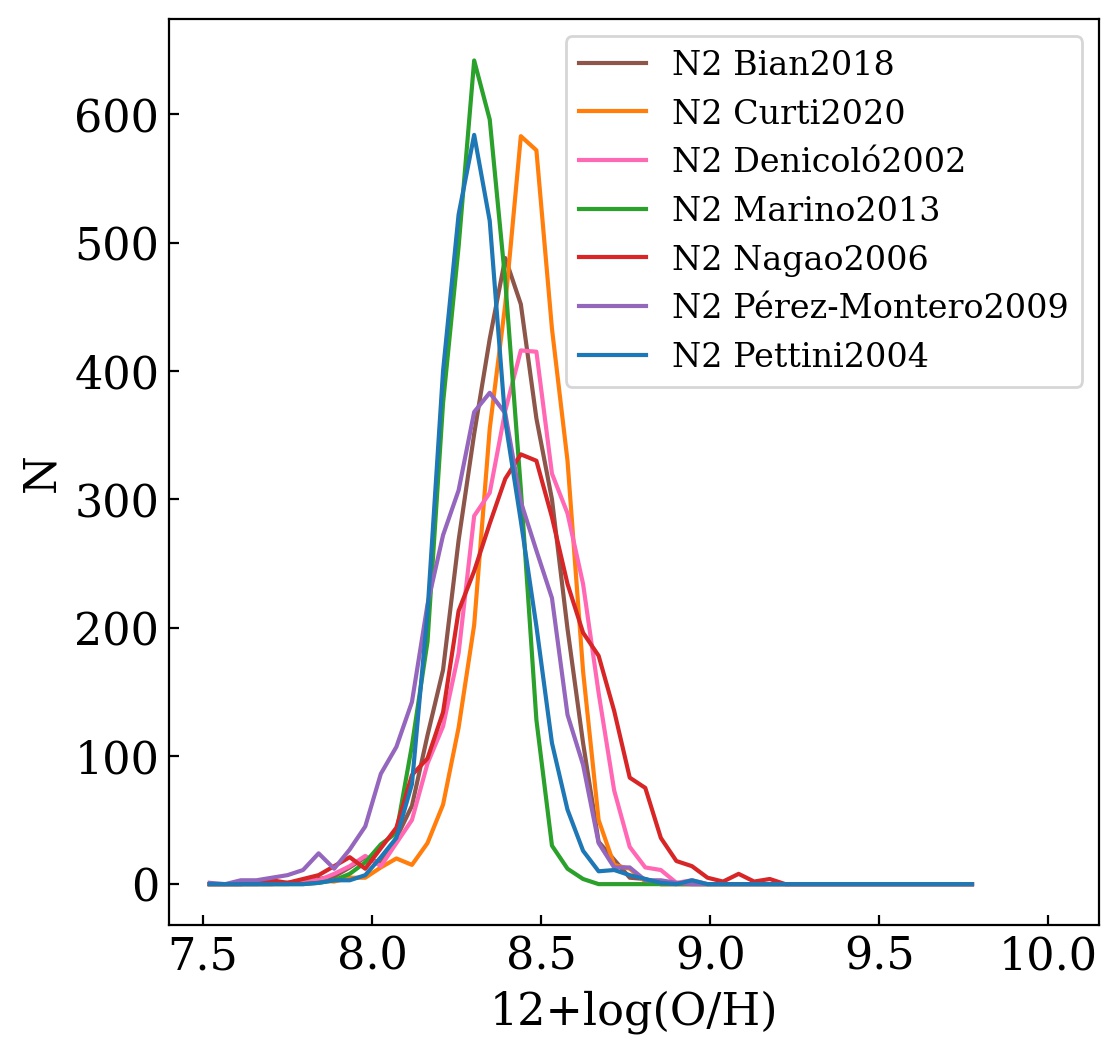}
    \includegraphics[scale=.41]{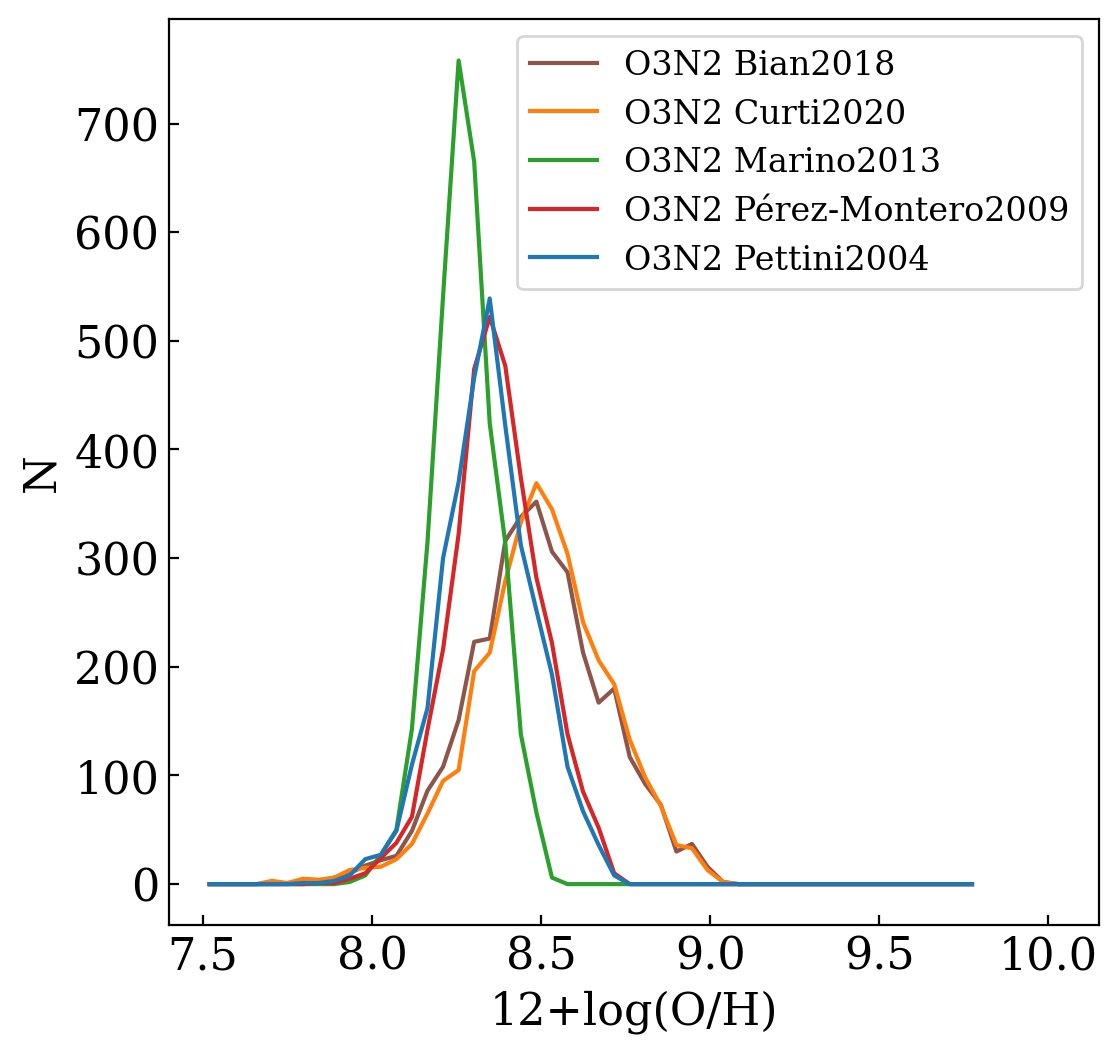}
    \includegraphics[scale=.41]{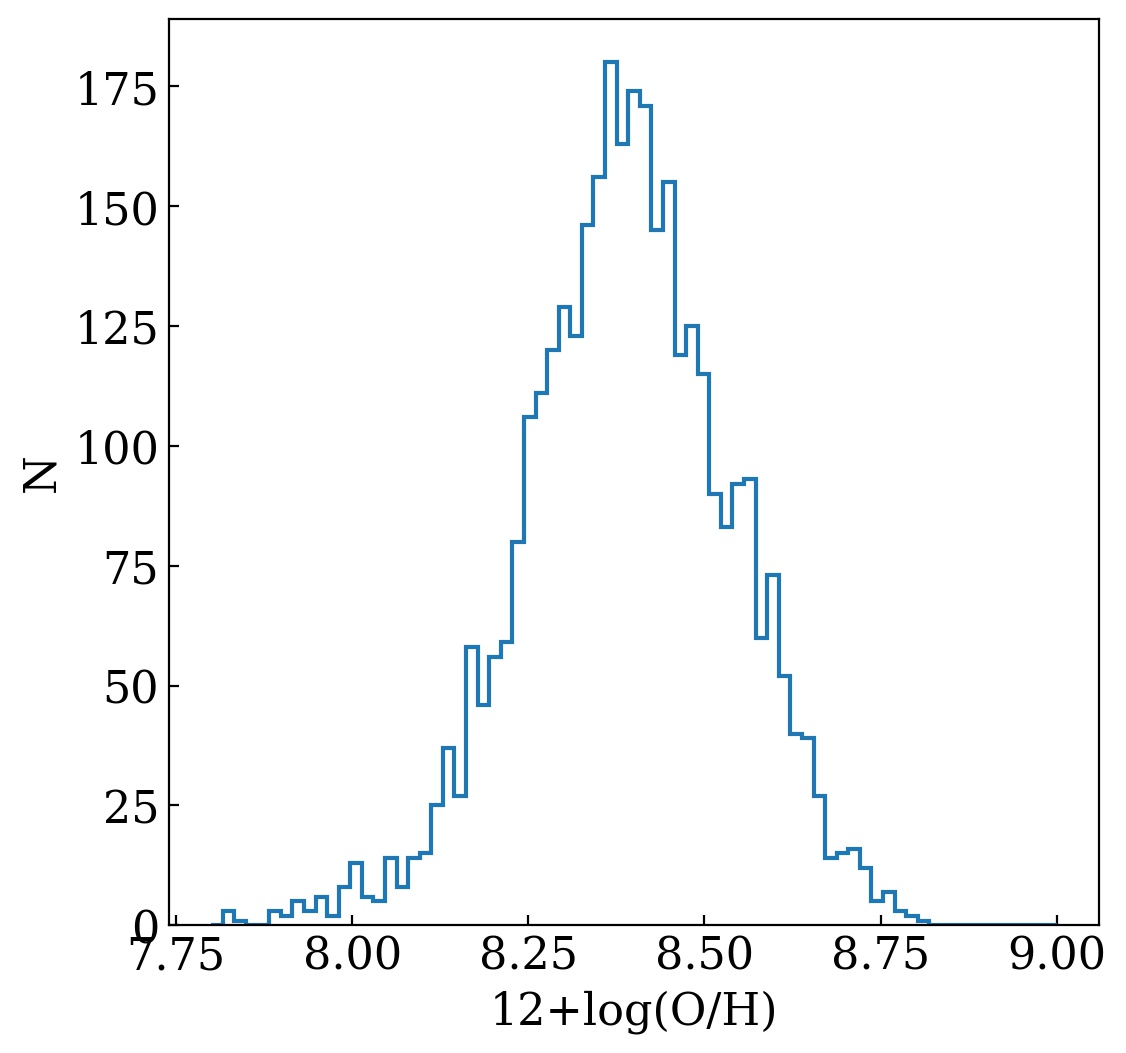}

    \caption{Metallicity distribution for our \textcolor{black}{dwarf galaxy catalogue} using N2 (left panel), O3N2 (middle panel) calibrators and the average of all these calibrators, listed in  Table \ref{tab:met} (right panel).}
    \label{fig:met}
\end{figure*}

Figure \ref{fig:met} shows the metallicity distribution for our dwarf galaxy catalogue for all diagnostics used in our study, i.e., N2 (left panel), O3N2 (middle panel) and the average of all calibrators used in this study (right panel). In Table \ref{tab:met} we show the mean metallicity  and standard deviation for our  catalogued dwarf galaxies  for each of the methods described above.
\textcolor{black}{It is well known that  N2 depends on the ionization parameter. O3N2 also depends on the ionization parameter, but to a lesser extent than N2.  According to \cite{Zurita21}, O3N2 overestimates gas--phase metallicities for low--metallicity galaxies. However, the dwarf catalogue galaxies have near solar metallicities. }  We compared the mean  12 + log(O/H) for the dwarfs in our catalogue derived from O3N2+N2 and O3N2 and N2 calibrators and found no  difference between them within the uncertainties, see Table \ref{table2}.

  As our catalogue contains dwarfs with a range of properties no particular calibrator is ideal for all cases, so for our analysis we utilise the 12 + log(O/H) for each catalogued dwarf's  abundance using  the average of the metallicity from all of O3N2 and N2 calibrator methods listed in Table \ref{tab:met}. 
 Appendix \ref{comparison} contains a comparison of our metallicity results with those produced by MPA--JHU. From that comparison, we can say that our metallicity estimates show a similar trend to MPA--JHU if we use the same strong line calibrations.

\begin{table} 
\centering
\caption{\textcolor{black}{Mean metallicity for the dwarf galaxy catalogue from the N2 and O3N2--based calibrators used in this study.}} \label{tab:met}
\begin{tabular}{llcccccc}
    \hline 
Calibrator&Reference         &  Mean   &  standard deviation \\
    \hline
N2& Bian2018       &  8.39  &  0.14 \\   
N2 &Curti2020      &  8.45  &  0.12 \\
N2& Denicoló2002   &  8.44  &  0.16 \\
N2& Marino2013     &  8.31  &  0.10 \\
N2 &Nagao2006      &  8.45  &  0.20 \\
N2& Pérez--Montero2009  &  8.33  &  0.18 \\
N2& Pettini2004    &  8.33  &  0.12 \\
O3N2& Bian2018     &  8.49  &  0.20 \\

O3N2& Curti2020   &  8.51  &  0.20 \\
O3N2&  Marino2013    &  8.28  &  0.09 \\
O3N2& Pérez--Montero2009  &  8.37  &  0.13 \\
O3N2& Pettini2004  &  8.35  &  0.12 \\
    \hline
\end{tabular}\\
\end{table}

\begin{figure}
	\includegraphics[scale=.57]{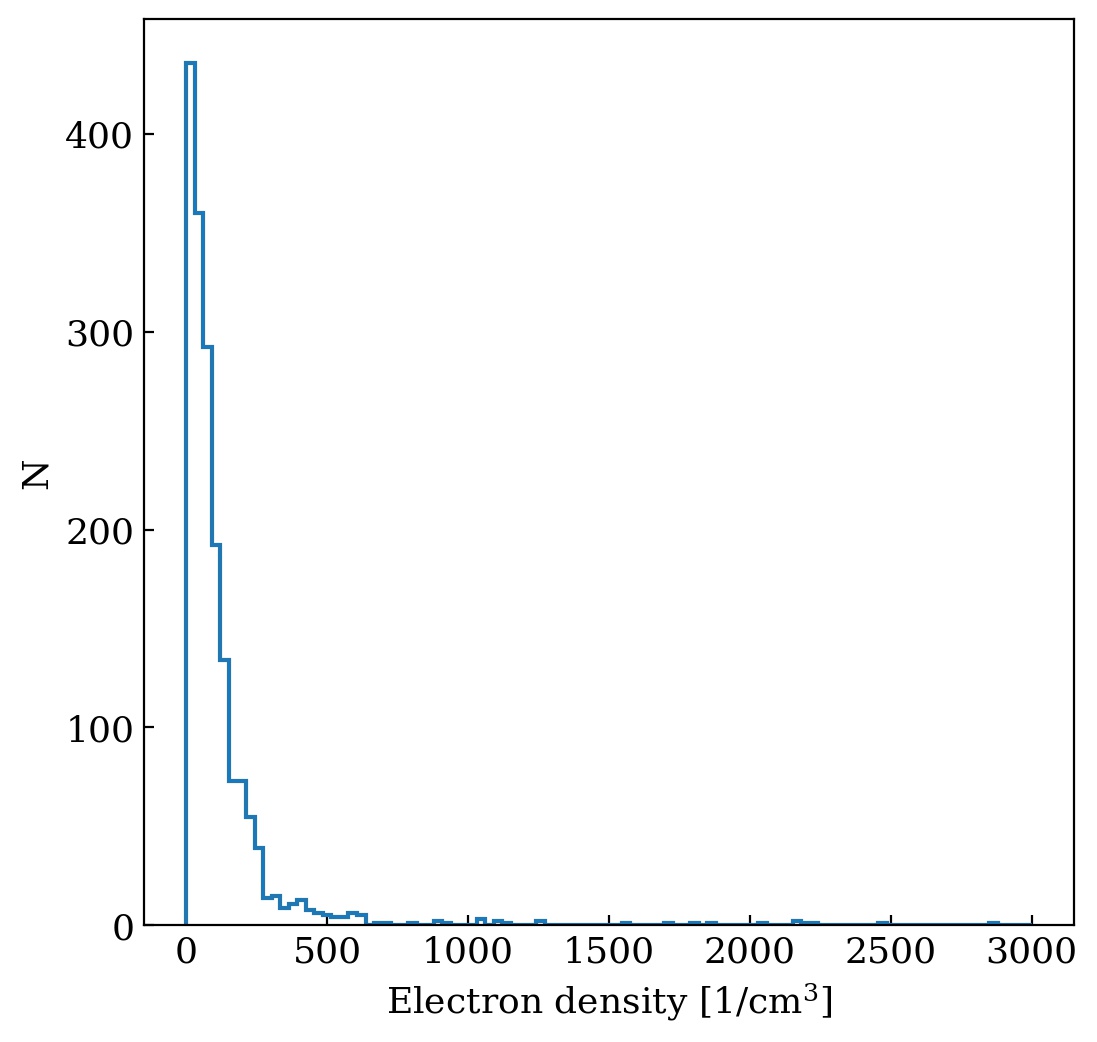}
    \caption{The distribution of electron densities for the dwarfs in our catalogue with non-saturated electron densities.}
    \label{fig:density}
\end{figure}

Additionally, we  calculated the electron density n$_e$  from  the emission line ratio [SII]$\lambda$6717/[SII]$\lambda$6731, assuming 
T$_e$(OIII) = 10000 K,  using \textsc{temden} implemented within the \textsc{IRAF}\footnote{Image Reduction and Analysis Facility (IRAF) is distributed by the National Optical Astronomy Observatories,
which are operated by the Association of Universities for Research
in Astronomy, Inc., under cooperative agreement with the National
Science Foundation.} \textsc{nebular} software package. Figure \ref{fig:density} shows the resulting distribution of electron densities for the catalogued dwarfs with non-saturated electron densities.
\textcolor{black}{We found that only 1780 galaxies, in our  catalogue, have non--saturated electron density values ([SII]$\lambda$6717/[SII]$\lambda$6731<1.43). In that sub--sample most galaxies show low electron density values of the order of $\sim$100 cm$^{-3}$ which is normal for star--forming dwarf galaxies \citep[e.g.,][]{Masegosa1994,Bordalo2011}.}

\subsection{Star formation rate determination}
\label{result:sfr}

For each galaxy in  the catalogue, we estimate the SFR from it's H$\alpha$ luminosity L(H$\alpha$) using the \cite{kenn98} conversion, \textcolor{black}{which assumes a  Salpeter IMF and solar metallicity. To ensure consistency amongst our measurements we then made a conversion} to a Chabrier \citep{Chabrier2003} IMF, i.e., SFR(H$\alpha$) (M$_{\odot}$ yr$^{-1}$) =  L(H$\alpha$)/10$^{41.31}$ erg s$^{-1}$ \citep{Miranda2023}. We compared our SFR estimates with the MPA--JHU catalogue  in Appendix \ref{comparison}.
The fluxes in the MPA--JHU catalogue  are normalised to  match the photometric r -- band fibre magnitude. This normalisation factor was derived from the MPA--JHU catalogue and was applied to the  galaxies in our catalogue. We also converted the MPA--JHU SFR(H$\alpha$) (Kroupa IMF) to Chabrier IMF by decreasing the MPA--JHU SFR (Kroupa IMF) by 0.03 dex \citep{Miranda2023} to enable comparison with our SFR(H$\alpha$) (See Appendix \ref{comparison}). The 3$^{\prime\prime}$ diameter SDSS fibres cover only a small fraction of our dwarfs' disk areas, see Section \ref{em_line_flux}. So, the SFR(H$\alpha$) estimated from the SDSS fibre data alone would in almost most cases underestimate the SFR(H$\alpha$) of our catalogued dwarfs. Thus as a final step, to estimate the total SFR(H$\alpha$) of the catalogue dwarfs, we used the aperture correction procedure prescribed in Section 4.1 of \cite{Miranda2023}. We use the aperture corrected SFR(H$\alpha$) values of our dwarfs for our analysis throughout the paper and these values are presented in the online catalogue and Table \ref{table3}. The only exception is the comparison with MPA--JHU in Appendix \ref{comparison} which is a comparison of SFR(H$\alpha$)s before aperture correction.

\subsection{Dwarf galaxy catalogue}
\label{results:catalogue}

 Our final dwarf catalogue contains   3459 galaxies  and is available online  (see Appendix \ref{online}).  Table \ref{table3} shows a truncated version of the catalogue with example entries. The columns in the Table are as follows:\\
 (1) The SDSS SpecObject ID\\
 (2 and 3) RA DEC J2000 positions\\  
 (4) SDSS galaxy name\\
 (5) SDSS spectroscopic redshift\\
 (6) g -- i SDSS model magnitude colour\\ 
 (7) D$_{25}$ [kpc]\\
 (8) log(M$_*$) [\msolar]\\
 (9) log(SFR(H$\alpha$)) [\msolaryr]\\
 (10) EW(H$\beta$) [\angs]\\
 (11) 12 + log(O/H) and 1$\sigma$ uncertainty. Mean of values for each  dwarf is derived from   all N2 and O3N2 calibrators listed in Table \ref{tab:met}.\\
  (12) SDSS photometery flag, 0 = not clean, 1 = clean, 2 = derived M$_*$ or g -- i colour is unrealistic. 
 
  The  majority of our catalogued dwarfs have stellar masses between $\sim$10$^{\textcolor{black}{8}}$--10$^{9}$ M$_{\odot}$ (see Figure \ref{fig:result}), which are normal values for dwarf galaxies. Although the catalogue does contain a small fraction of low stellar mass dwarfs with stellar masses $<$10$^7$ M$_{\odot}$. Figure \ref{fig:met} (right panel) shows the distribution of the mean 12 + log(O/H) for the galaxies in the catalogue. The catalogue's mean 12 + log(O/H) value is  $\sim$8.39$\pm$0.15. This mean value \textcolor{black}{falls toward the upper end of the range of expected  12 + log(O/H) of 7.4 to $\lesssim$ 8.4 referred to in Section \ref{intro}. The absence of the lowest metallicity XMP dwarfs} is likely the result of applying the 2.5 $\sigma$ line emission SNR  selection criterion to the catalogue. No correlation between the catalogue's 12 + log(O/H) and redshift was found. For further detail about metallicity and its calibration, see Section \ref{result:metal}.
 
 The distributions of D$_{pet}$  [kpc],  SDSS g -- i colour,  log(M$_*$/\msolar) and redshift for the galaxies in the catalogue are shown in Figure \ref{fig:result}. The top  left panel of  Figure \ref{fig:result} shows the  catalogue's dwarfs have mean  D$_{pet}$ = 3.66$\pm$1.33 kpc, (D$_{25}$ = 5.48$\pm$1.99 kpc). This confirms their D$_{25}$ diameters lie within the $\sim$ 0.1 kpc -- 10 kpc  range expected for dwarf galaxies, in Section \ref{intro}. Table \ref{table2} gives the catalogue's mean values and  the standard deviations for  12 + log(O/H), EW(H$\beta$),  SDSS g -- band magnitude,  D$_{25}$ g -- band  diameter,   reshift, g --i colour, log(M$_*$) and log(SFR(\halpha)) \textcolor{blue}{(total SFR)} calculated  as described in  Section \ref{result:sfr}.

\begin{table}
\footnotesize
\begin{minipage}{120mm}
\caption{ Dwarf galaxy  catalogue properties}
\label{table2}
\begin{tabular}{@{}lcll@{}}
\hline
Property&Mean& standard& Unit \\
 & & deviation & \\
 \hline
12 + log(O/H) O3N2+N2& 8.39 & 0.15&\\
 12 + log(O/H) N2&  8.38 & 0.15&\\
 12 + log(O/H) O3N2&  8.39 & 0.13&\\

EW(H$\beta$) &\textcolor{black}{18.80} & 24.27 & ${\textrm{\AA}}$\\

g -- band magnitude & -16.53 & 1.35 & mag \\
g -- band \textcolor{black}{D$_{25}$}& 5.48& 1.99 &kpc\\
Redshift& 0.017327& 0.00816&\\
g -- i colour&\textcolor{black}{0.52}& 0.47 & mag\\
	log(M$_*$) &\textcolor{black}{8.43}& 0.74& \msolar\\
	log(SFR)& \textcolor{black}{-1.32}& 0.67 & \msolaryr\\
 \hline
\end{tabular}
\end{minipage}
\end{table}


\begin{table*}
\footnotesize
\begin{minipage}{\linewidth}
\caption{\textcolor{black}{Dwarf galaxy catalogue -- example}}
\label{table3}
\begin{tabular}{@{}lllllrrrrllllllll@{}}
\hline
SpecObjID&RA&Dec&SDSSname&Redshift&g--i&D$_{25}$&log(M$_*$)&log(SFR)&EW(H$\beta$)&12+log(O/H)&Flag\\
&[hms]&[dms]&&&&[kpc]&[\msolar]&[\msolaryr]&[\angs]\\
1 \footnote{SpecObjID is the SDSS identification number unique to each pointing. Due to lack of space the SpecObjID has been shortened and represented as Galaxy1, Galaxy2 etc in the first column. The complete IDs are the following: 299496824270514176, 299582311299573760, 299619694694918144, 299647182485612544, respectively. The online version of Table 3 contains the complete SpecObjectIDs in column 1. }&2&3&4&5&6&7&8&9&10&11&12\\
\hline
Galaxy1 &	9h48m42.33s&	-0d21m14.50s&	SDSS J094842.33-002114.5&	0.00628	&0.76&	4.2&	8.20&	-1.87&	4.89&	8.37$\pm$0.08&1\\
Galaxy2 &	9h41m17.02s&	+0d46m16.00s&	SDSS J094117.02+004615.9&0.00658	&0.69	&1.9&	7.88& -1.90	&6.98	&8.42$\pm$0.09&1\\
Galaxy3 &	9h44m10.90s	&+0d10m47.17s&	SDSS J094410.90+001047.1&	0.01112	&0.73& 5.0 &8.67	&-1.09	&65.71	&8.28$\pm$0.05&1\\
Galaxy4 &	9h47m05.50s	&+0d57m51.24s	&SDSS J094705.50+005751.3	&0.00618	&1.00&	6.3	&8.58	&-2.59	&19.63&	8.28$\pm$0.06&0\\

 \hline
\end{tabular}
\end{minipage}

\end{table*}

As expected, the stellar mass distribution in the catalogue (Figure \ref{fig:result}) strongly favours dwarfs with masses near our  catalogue's M$_*$ upper limit (1 $\times\ $ 10$^9$\msolar), because such dwarfs being more luminous have a much higher probability of having a SDSS spectral observation than less luminous dwarfs. Moreover, more massive star--forming dwarfs have a higher probability of fulfilling the our emission line SNR criteria.  The catalogue's dwarfs have  a mean  M$_*$ = 2.7 $\, \times$ 10$^8$ \msolar, i.e., similar to the Small Magellanic cloud,  M$_*$ = 3.1 $\times $ 10$^8$ \msolar\  \citep[][]{vandermarel09}.
 
A g -- i colour of 1.1 approximately coincides with the red--sequence threshold displayed in the colour-–magnitude plot in ﬁg. 12 of\,  \cite{cort08}. The galaxies in our catalogue  have a mean SDSS g -- i\, colour of  0.52$\pm$0.47 (top right panel of Figure \ref{fig:result}). \textcolor{black}{On average the galaxies in the catalogue have} a g -- i colour $<$ 1.1,  unambiguously qualifying them  as blue galaxies.   This confirms that the catalogue is dominated by dwarfs which have experienced a recent episode of SF.  We note that our catalogued dwarfs do not display the bimodal colour distribution seen in some other samples, e.g.,  the low surface brightness galaxies in  \citep{Bhattacharyya23}. The absence of dwarfs with red colour in the catalogue is likely to be the result of the our selection bias, see section \ref{mass_metal}.  
\textcolor{black}{We see from the bottom right panel of Figure \ref{fig:result} that the catalogue's redshift distribution has several maxima with two of  the most prominent at  z $\sim$ 0.006 and z $\sim$ 0.023. The number of  dwarfs between z $\sim$ 0.025 and 0.03 declines rapidly, reflecting the rapid fall off of completeness of SDSS dwarf galaxies meeting our criteria beyond z $\sim$ 0.025. The redshift and spatial distribution of the catalogue's dwarf's are discussed further in section \ref{discus:env}.}  

\section{Discussion}
\label{discuss}

\subsection{Mass--metallicity/SFR relations
\label{mass_metal}}
 This subsection investigates the stellar mass gas--metallicity relation of our catalogue, within a stellar mass range of log(M$_*$/\msolar) $\sim$ 6 to 9,  in comparison to other samples.  This relation has been reported for massive galaxies by several authors \cite[e.g.,][]{tremonti04,gallazzi05} and for dwarf galaxies in a similar mass range to our catalogue \cite[e.g.,][]{jimmy15}.
As M$_*$ values in the relation depend on robust photometry, for this analysis we  used a sub--sample of our catalogue (n = 2568) restricted to galaxies with a clean photometry flag = 1, in Table \ref{table3}.

\begin{figure*}
\includegraphics[scale=.6]{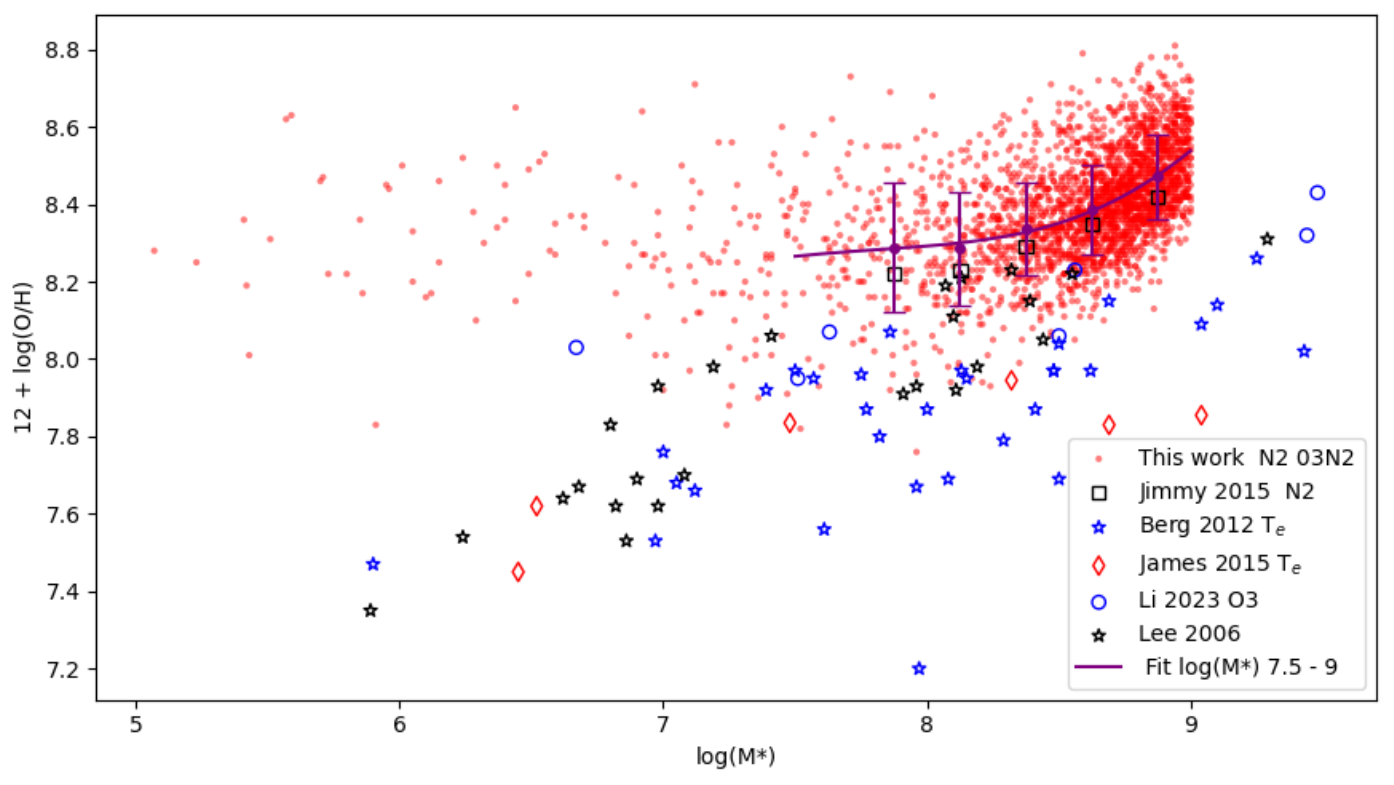}
\caption{Mass metallicity relation, 12 + log(O/H) v log(M$_*$/\msolar),  for the sub--sample of dwarfs in our catalogue with clean SDSS photometry compared to values from the literature (see the text). A spline (k=5) fits to  12 + log(O/H) v log(M$_*$/\msolar) for our sub--sample  for dwarfs, in the  log(M$_*$/\msolar) range 7.5 to 9, is shown with a magenta line.  Additionally, magenta dots show the mean with 1 $\sigma$ uncertainty error bars for our sub-sample's 12 + log(O/H) values, in the same mass bins used by \citet{jimmy15}.} 
    \label{fig:mass_metal}
\end{figure*}

In Figure \ref{fig:mass_metal}, we compare the values from our sub--sample with those from \cite{lee06}, \cite{berg12}, \cite{James15}, \cite{jimmy15} and \cite{li23}.  In the same figure, we show a spline--fit (K=5) to our data for dwarfs in the stellar mass range log(${M}_{* }$/\msolar) = 7.5  to 9. We see from the fit to our data that  our 12 + log(O/H) values, based on the mean of O3N2 and N2 calibrators, are in good agreement, within the uncertainties, with those from \cite{jimmy15} in the mass range log(${M}_{* }$/\msolar) $\sim$ 8 to 9. However, our 12 + log(O/H) values are systematically higher for the same stellar mass compared to those from \cite{lee06}, \cite{berg12}  and \cite{James15}. 

The plot also shows values from higher redshift (z = 2 to 3) dwarfs from \cite{li23}.  The slope of their relation becomes shallower at log(M$_*$/\msolar) $\lessapprox$ 8.5,  similar to trends seen in  our data and \cite{jimmy15}. In contrast, the Berg et al., James et al. and Lee et al. data imply the gradient of the relation found at log(${M}_{* }$/\msolar)  $\gtrapprox$ 8.5 remains constant down to stellar masses as low as log(M$_*$/\msolar) $\sim$ 6. 

 The Berg et al. and James et al. 12 + log(O/H) values are derived from long--slit Multiple Mirror Telescope (MMT) spectra which could potentially be  more representative of the metallicity of the whole galaxy compared to the more limited sampling at the centre of dwarfs by the SDSS fibre aperture for our catalogue. However, the expected rapid dispersion of metals in dwarfs \citep[e.g.][]{Kobulnicky1996, croxall09, lagos2009,lagos2016} argues against this as an explanation for the large observed differences between the samples. Because of the limited SDSS spectral coverage,  \cite{jimmy15} uses N2 and we use N2 and O3N2 calibrators rather than the preferred direct method (T$_e$) to determine 12 + log(O/H). 
 However, the 12 + log(O/H)  N2 and O3N2 calibrators used in this study are formulated to give similar results to the T$_e$ method. It is therefore highly unlikely that the adopted calibrator(s) alone are responsible for the  differences  in mass--metallicity slopes  in different samples observed in Figure \ref{fig:mass_metal}.

 To analyse and ascertain the exact reasons for the differences in the results in Figure \ref{fig:mass_metal} is beyond the scope of this paper.  The [OIII]$\lambda$4363 line is faint in sub--solar metallicity where metal cooling is more efficient. This lead us to use strong--line calibrators instead of the T$_e$--based method to calculate the oxygen abundance.  However,  we consider our emission line SNR criterion to be the principal reason why the mean metallicity in each mass bin shows a systematically higher metallicity compared to the other samples in Figure \ref{fig:mass_metal}. 
 Our SNR criterion led to the exclusion of metal--poor dwarfs with weak emission lines from our calculation of the mean metallicity of each mass bin. This had the effect of raising the calculated mean metallicity of each mass bin compared to the underlying metallicity of the bin's dwarfs.

 Additionally, there is a trend for emission lines to become proportionately weaker  with decreasing stellar mass.  Therefore, as  stellar mass decreases  the fraction of dwarfs  able to meet our SNR criterion also decreased. This declining fraction meant the reported mean metallicities of successively lower mass bins were based on  increasingly unrepresentative samples of high--metallicity dwarfs and an under representation of the bin’s underlying and predominantly low--metallicity dwarf population. 
 We, therefore, consider this progressive under representation of the low--mass low--metallicity population is likely to be responsible for the apparent flattening of mean metallicity derived for the mass bins, below log(M$_*$/\msolar) $\sim$ 8.5 in Figure \ref{fig:mass_metal}.

 Using chemical evolution models \cite{Recchi2015} investigated the chemical evolution of ancient TDGs formed out of metal--poor baryonic tidal interaction debris at high redshifts. They found that such ancient TDGs formed out of gas with initially very low metallicity, naturally increase their metallicity  to follow the linear mass--metallicity relation for local dwarf galaxies from  \cite{lee06}, see Figure \ref{fig:mass_metal} and Figure 1 in \cite{Recchi2015}. In contrast TDGs formed at lower redshifts from metal--enriched baryonic debris are observed to be outliers above the mass--metallicity relation at log(M$_*$/\msolar) $\lessapprox$ 9, see figure 1 in \cite{Recchi2015}. Interestingly,  the shallower mass--metallicity relation for our catalogue is similar to the one reported by \cite{Recchi2015} for more recently formed high--metallicity TDGs \citep{boquien10,duc14}.  We speculate this similarity may indicate that our catalogue could contain a significant number of TDGs formed at lower redshifts. However, this seems highly unlikely to be the explanation for the bulk of catalogue members.

\begin{figure}
\includegraphics[scale=.42]{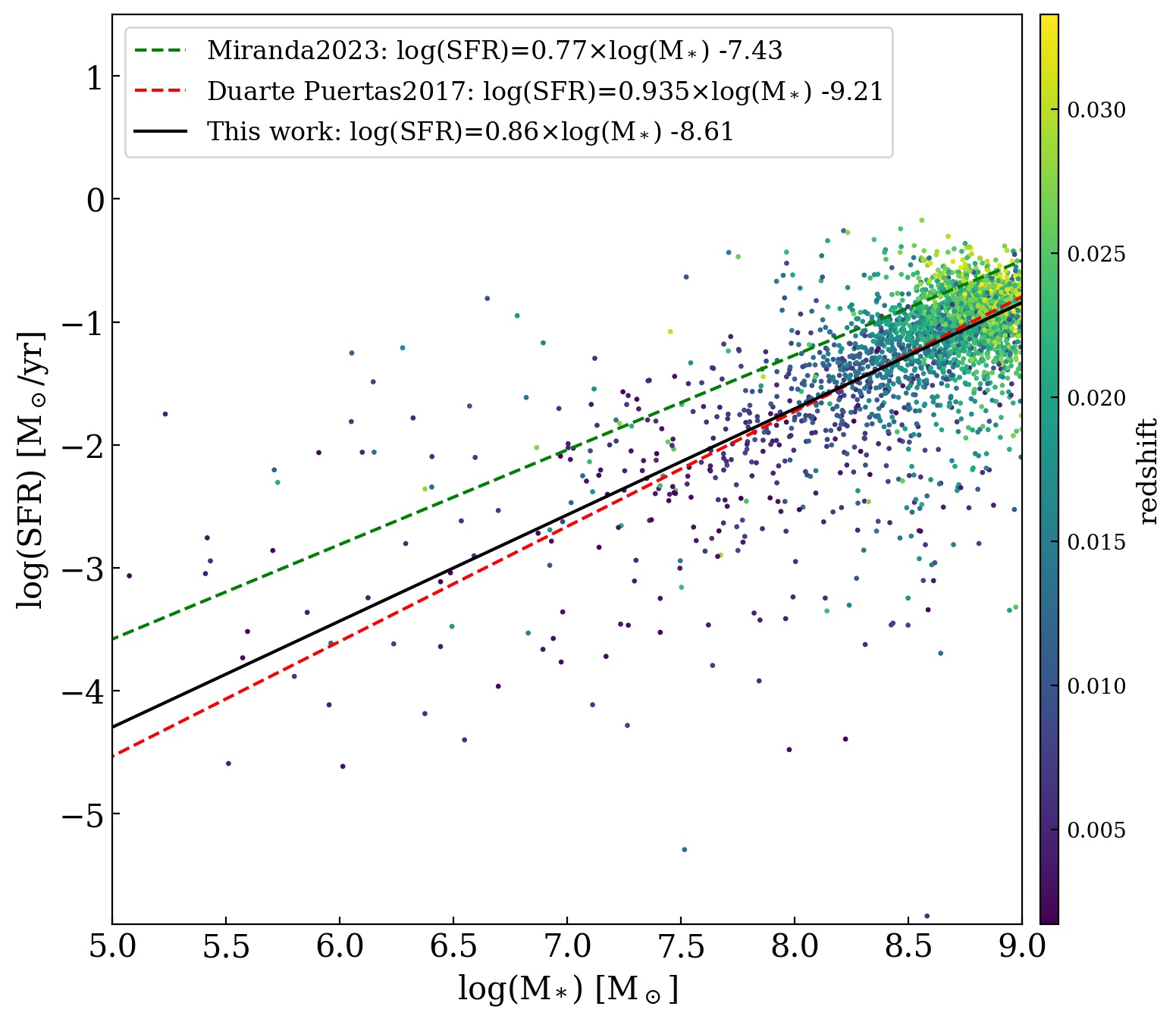} 
\caption{A plot of log(SFR/\msolaryr) against log(M$_*$/\msolar) for  the sub--sample of dwarfs in our catalogue with clean SDSS photometry. The linear fit to our sub--sample data is shown with  black  line.  To compare to our fit to the literature, we also show  linear fits from \citet{Duarte17} --- red dashed line,  and \citet{Miranda2023} -- green dashed line. See text for comments on the slope differences.}
\label{fig:sfr}
\end{figure}

 Figure \ref{fig:sfr} shows the log(SFR/\msolaryr) vs log(M$_*$/\msolar)  for the catalogue. Since  this relation requires M$_*$, the Figure uses same sub--sample we used for mass--metallicity analysis. The Figure also shows a linear fit to our sub--sample data.  Additionally, the plot shows the linear fits to the relations from \cite{Duarte17} and \cite{Miranda2023} for the SDSS galaxies in their samples.  The slope of the fit to our data (black line) is in good agreement with the one from \cite{Duarte17} (red line) which is derived from SDSS spectra with  aperture corrections based on the CALIFA IFU survey observations. This agreement confirms that our aperture corrections for the clean photometry subsample are reasonable. However, the slope of the fit to our sub--sample is slightly steeper and offset  compared to the fit to Miranda's SDSS sample (green line), which is dominated by main sequence star--forming galaxies with log(M$_*$/\msolar) $>$ 9.

\subsection{Environment }
\label{discus:env}
\begin{figure*}
	\includegraphics[scale=.49]{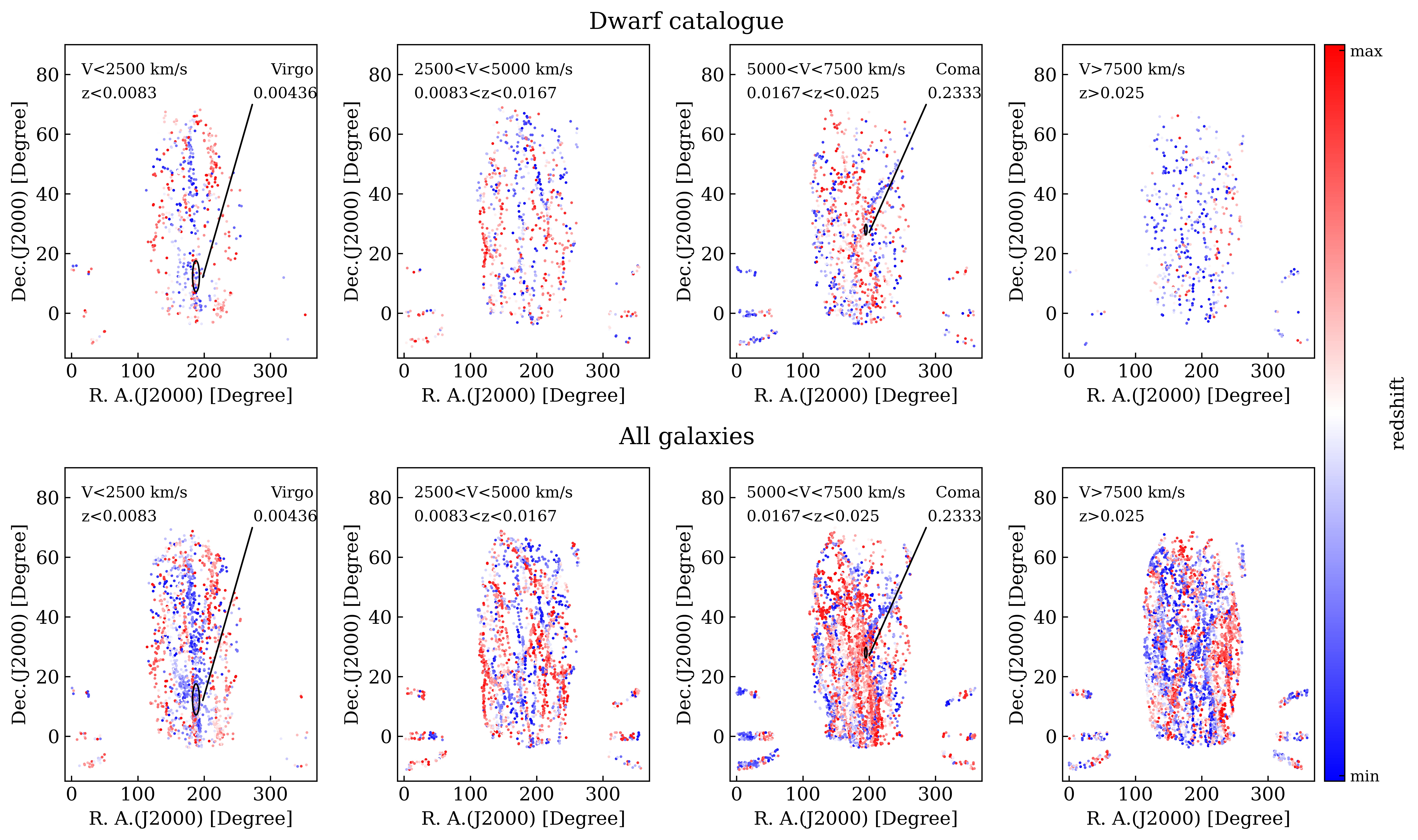}
    \caption{\textcolor{black}{Sky distribution of the SDSS--DR16 galaxies with spectra in 4 redshift bins (columns): The top row shows the catalogued dwarf galaxies and the bottom row all non--dwarf galaxies with SDSS spectra.  The relative redshifts of  galaxies within each redshift bin are indicated with} the colour per the colour scale.  The Virgo and Coma galaxy clusters' sky positions and sizes are shown on the Figure with the  larger and smaller ellipses, respectively. }
    \label{fig:3a}
\end{figure*}

The redshift distribution of the catalogue's dwarfs, Figure \ref{fig:result} bottom right panel,  shows multiple maxima, with  the vertical lines on the plot  indicating the redshifts of the Virgo and Coma clusters. Two of the catalogue's redshift maxima are near, but not coincident with, the redshifts of Virgo and Coma.    Figure \ref{fig:3a} plots the sky positions of the dwarf (top row) and non--dwarf galaxies (bottom row) with SDSS spectra, each divided into four redshift bins. The colour of the points corresponds to the relative redshift of the galaxies within each redshift bin, with blue points representing lower redshift galaxies and red points, those with higher redshifts. The sizes of the Virgo and Coma clusters based on their respective virial radii (r$_{vir}$) of  6\degree\ or 1.7 Mpc \citep[][]{mei07} and  $\sim$ 1.5\degree\ $\sim$ 3 Mpc \citep{lokas03} are shown in the Figure with ellipses.  It is clear from Figure \ref{fig:3a} that the dwarfs in our catalogue are not preferentially associated with the Virgo and Coma galaxy clusters as might be suggested by the redshift distribution in Figure \ref{fig:result} -- bottom right panel. Instead, they are distributed across much larger redshift and spatial scales.  For  dwarfs  with  velocities below 7500 \km\  their distribution in RA, DEC and redshift space trace the same filamentary structures  traced by the SDSS non--dwarf galaxy population with spectra. At velocities above 7500 \km\ (z = 0.025) the association between the dwarf and non--dwarf galaxies  remains  only for the lowest redshift dwarfs.  For dwarfs with higher redshifts in this bin, the association between the populations  progressively dissolves with increased redshift,  with few dwarfs  meeting our selection criteria at the highest redshifts. We conclude  that at velocities below $\sim$ 5000 \km\ the catalogued dwarfs trace lower--density large--scale galaxy structures reasonably well, but  above 5000 \km\, the association between dwarf and non--dwarf galaxies  increasingly weakens, due to the relative incompleteness of the dwarfs in our catalogue.

 To assess the effect of the local environment on the catalogue's properties, we carried out an analysis of the impact of tidal forces within a limited volume surrounding each catalogued dwarf and the tidal forces acting on the dwarf arising from its nearest neighbour, as follows: 

 \cite{byrd90} investigated the triggering of nuclear SF from tidal encounters in Seyfert galaxies using a tidal perturbation parameter, $P_{gg}$, (Equation \ref{eqn1}):

\begin{equation} \label{eqn1}
	P_{gg} = \frac{(M_c/M_g)}{(r/A)^3}  
	\end{equation}
	
 where $M_c$ =  M$_*$ of the nearest companion galaxy, $M_g$ =  M$_*$ of the target galaxy, r =  projected distance between the two galaxies and A = is  the optical disk radius of the target.  While the threshold for triggering SF depends on the specifics of the interaction, our analysis assumes the tidally induced star--formation threshold is in the $P_{gg}$ range of  0.006 -- 0.1 from \cite{byrd90}.

To define the environment of isolated galaxies \citep{verley07} developed a tidal force parameter, which is a measure of the sum of tidal forces acting on a galaxy in comparison to the binding force within it. In our analysis, we modified Verley's tidal force parameter to include only companions with SDSS  spectroscopic redshifts within a radius of 500 kpc and velocities within $\pm$ 300 \km\ of the target galaxy. Our tidal force parameter (Q$_{0.5-300}$) takes into account both the projected separation distance and the M$_*$ of the companions,  calculated as follows:
{\bf
\begin{equation} \label{eqn2}
	Q_{0.5-300} = log(\sum_{i} Q_{ig})  
	\end{equation}}

{\bf \begin{equation} \label{eqn3}
	Q_{ig} = (M_i/M_g)(D_g/S_{ig})^3  
	\end{equation}}
	
 where $M_i$ = the companion's M$_*$, $M_g$ = the target's M$_*$, D$_g$ is the target's D$_{25}$ and S$_{ig}$ is the projected distance between the companion and the target galaxy. Values used for these calculations were extracted from the SDSS online database, with D$_{25}$ = 1.5D$_{Petrosian}$ and M$_*$,   calculated using the SDSS g and i model magnitudes together with coefficients  from \cite{bell03}. The Q$_{0.5-300}$ parameter  focuses on companions which could potentially have interacted with the target galaxy within the last $\sim$ 1 Gyr, so could have conceivably triggered  SF in the target galaxy. Because our analysis is limited to galaxies with SDSS redshifts, we are excluding companions which lack SDSS redshifts, and the impact of this limitation increases with redshift.  


\begin{figure}
\begin{center}
\includegraphics[ angle=0,scale=0.62] {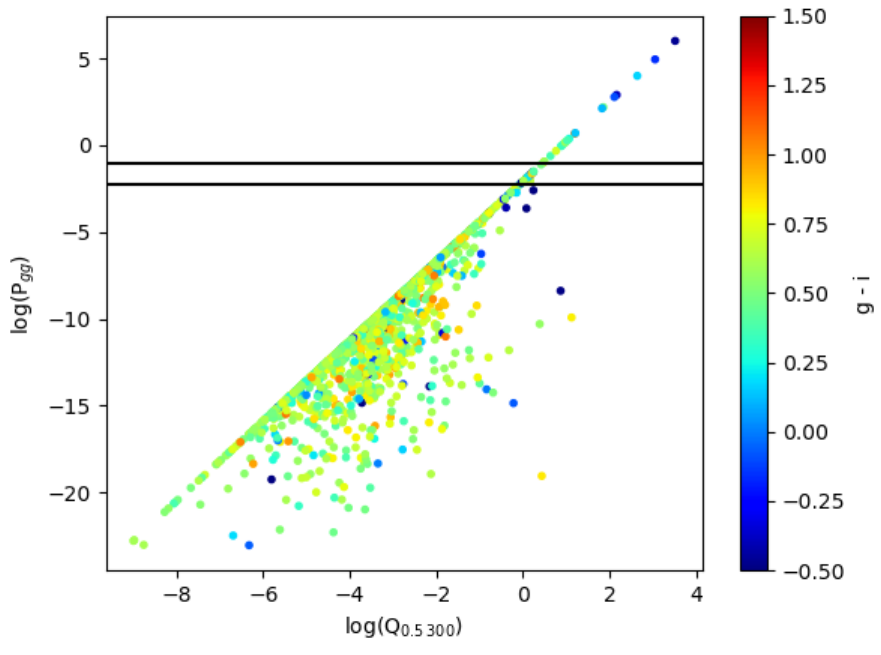}
\vspace{0.01cm}
\caption{ Nearest neighbour induced star formation parameter,  log($P_{gg}$), v log(Q$_{0.5-300}$) sum of near neighbour tidal forces parameter, for our sub--sample of dwarf galaxies with clean photometry flags in SDSS. The horizontal lines show the threshold range for tidally induced star formation  \citep{byrd90}. The g -- i colour of the galaxies is per the colour scale.}
\label{fig_12} 
\end{center}
\end{figure} 

 Figure \ref{fig_12} plots the tidal perturbation parameter P$_{gg}$    
against the  Q$_{0.5-300}$ parameter for our sample dwarfs with the  g -- i colour of the galaxies per the  colour scale.  The Figure plots the same clean photometry sub--sample (n = 2568) used in section \ref{mass_metal}. We used this sub--sample because P$_{gg}$ and Q$_{0.5-300}$ require parameters derived from SDSS photometry. Limiting our analysis to the clean photometry sub--sample improves the reliability of our analysis. The region between horizontal lines on the figure indicates the P$_{gg}$ threshold range for tidally induced star formation from \cite{byrd90}.  1112 of the sub--sample galaxies (43\%)  did not have a Q$_{0.5-300}$ companion. Only 96 sub--sample dwarfs (3.7\%) lie within or above the $P_{gg}$ threshold range for tidally induced SF. These  96 dwarfs are bluer than the vast bulk of the sub--sample which indicates the vast majority of the sub--sample is currently relatively isolated, with only a small fraction having a  P$_{gg}$ parameter and  colour consistent with having undergone a recent tidal interaction.

 log($P_{gg}$) and log(Q$_{0.5-300}$) values are based on the galaxies' current projected separation. It follows from the definition of Q$_{0.5-300}$ that all 1456 dwarfs in the plot could potentially have been close enough to their nearest companion to have crossed the tidally induced log($P_{gg}$) SF threshold within the past $\sim$ 1 Gyr. However, the 1456 dwarfs present a narrow range of blue colour 0.56$\pm$0.25. Moreover, leaving aside the 96 with the closest companions, the sub--sample lacks a  g -- i colour trend with either log($P_{gg}$) or log(Q$_{0.5-300}$), suggesting recent interactions are not driving the star-formation in sub--sample. Plots with the same log($P_{gg}$) and log(Q$_{0.5-300}$) axes as Figure \ref{fig_12} but using 12 + log(O/H) and log(SFR) for the colour axes, showed no trend between environment and 12 + log(O/H) or log(SFR).

\subsection{EW(H$\beta$) distribution and SFR}

\begin{figure*}
\includegraphics[scale=.49]{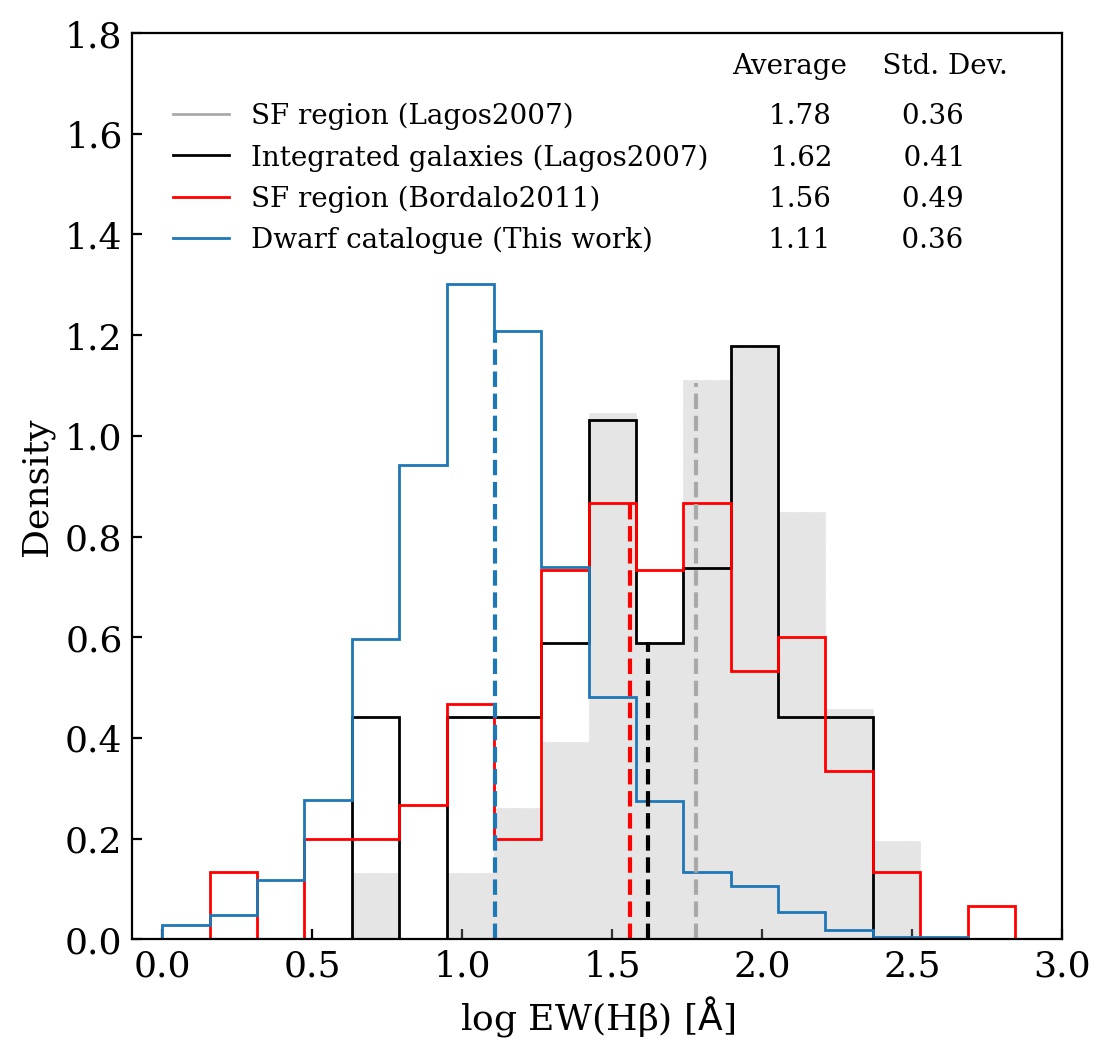}
\includegraphics[scale=.49]{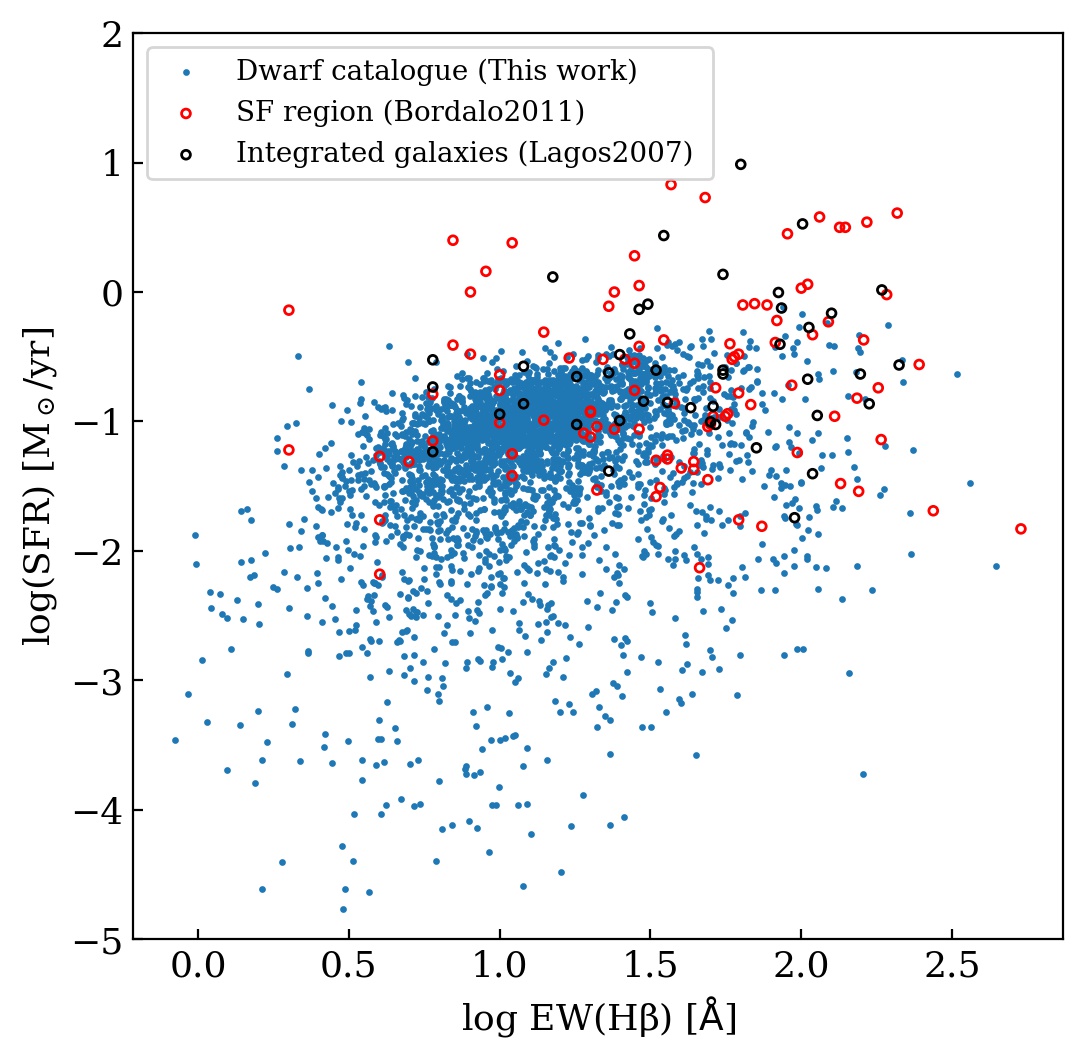}
    \caption{\textcolor{black}{EW(H$\beta$) distribution for our catalogue of galaxies \textcolor{black}{in comparison to distributions from the literature per the legend} (left panel). SFR versus EW(H$\beta$) \textcolor{black}{for our catalogue compared to literature samples per the legend} (right panel).}}
    \label{fig:ew-SF}
\end{figure*}

The EW(H$\beta$) distribution of our  catalogue is shown in Figure \ref{fig:ew-SF} (left panel).  This panel compares our results (mean log(EW(H$\beta$)/\angs) = 1.11) with those for strongly star--forming dwarfs (\hii/BCD galaxies) in the literature by \cite{lagos2007} and \cite{Bordalo2011} (mean log(EW(H$\beta$)/\angs) =1.56 and 1.78), respectively. EW(H$\beta$) has been used as an indicator of the age of the current burst of SF \citep[e.g.][]{Copetti1986}, with younger starbursts having higher EW(H$\beta$) values.  The EW(H$\beta$) distribution for our catalogue, spans a lower range of values compared to strongly star--forming  \hii/BCD galaxies, which indicates higher mean stellar ages for the current burst.

  In the right hand panel of Figure \ref{fig:ew-SF},  we see the  average log(SFR/\msolaryr) for the  HII/BCD galaxies is higher than that for our catalogue.We find a mean log(SFR/\msolaryr) of  -1.32, while the mean log(SFR/\msolaryr) of the HII/BCD galaxies are -0.59 for the \cite{Bordalo2011} sample and -0.57 for the \cite{lagos2007} sample. Interestingly, the mean log(SFR/\msolaryr) of our catalogue is between the mean log(SFR) of the HII/BCD galaxies and the mean value of gas rich low surface brightness (LSB) dwarfs (-2.04) \citep{McGaugh2017}.

 Gas infalls/interactions have been invoked to explain the SF activity in most \hii/BCD galaxies \citep[e.g.,][]{Pustilnik01,lagos2018}. But according to \cite{McGaugh2017}, the shallower  SFR--log(M$_{*}$) relation slope for dwarf LSBs,  compared to strongly star--forming \hii/BCD galaxies, can be explained without invoking gas accretion/interactions to sustain their SF. 
 This suggests that the moderate level of SF observed in \textcolor{blue}{most of} our sub--sample  dwarfs likely arises from stochastic SF bursts in relatively isolated dwarfs. This scenario is consistent with the finding in Section \ref{discus:env} that the overwhelming majority of the sub--sample dwarfs are relatively isolated and inhabit a rather narrow g -- i colour range.

\subsection{An overview (a possible scenario)}
 Recent work by \cite{cato23} showed that galaxies in filaments with the same redshift range as our catalogue display \hi\ deficiencies and  these deficiencies increase with proximity to the filaments \citep{Odekon18}. A reasonable explanation for these observations is  interactions between  galaxies in the filaments are removing \hi\ from the filament galaxies.  Section \ref{discus:env} shows the distribution of our catalogued dwarfs trace local galaxy filaments and it would be reasonable to expect dwarf galaxies in such filaments to be disproportionately impacted by intra--filament interactions, including interactions which generate high metallicity dwarfs, e.g., TDGs. From section \ref{mass_metal} we see that the our catalogued dwarfs have near solar metallicities comparable to TDGs formed within the last few Gyrs \citep{duc12}.   It would be interesting to investigate with future observations whether intra--filament interactions are capable of driving dwarf metallicities to the levels observed in our sub--sample particularly at masses below log(M$_*$/\msolar) =8.5. 

\section{Concluding remarks}

We present a catalogue of 3459 blue, nearby dwarf galaxies  (M$_*$ from 10$^{5}$ to 10$^9$ \msolar) extracted from SDSS DR--16   together with robust estimates of their metallicities. The strict spectral line SNR criterion used in our sample selection  biased our sample towards star--forming dwarf galaxies with bright emission lines, but it achieved the goal of creating a catalogue of nearby SDSS dwarf galaxies with robust metallicity estimates.

The catalogue was compiled as follows:  galaxies with radial velocity, 500 \km\ $\le$ velocity $\le$ 10000 \km (0.00166 $\le$ z $\le$  0.03333). To eliminate the massive galaxies, only sources with absolute magnitude M$_{g}$ $\ge$ -18 mag, stellar mass $\le$ 10$^{9}$ \msolar\ and an optical diameter (D$_{25}$) $\le$ 10 kpc were retained.
For a robust estimation of metallicities, we further limited our catalogue to galaxies where the emission lines used to determine the metallicities (H$\beta$ 4861\angs, [O III] 5007\angs, H$\alpha$ 6563\angs, [N II] 6584\angs, [S II] 6717\angs, [S II] 6731\angs) all had an SNR $\ge$ 2.5.  
\begin{enumerate}

\item
We compared the mass--metallicity relation from our clean photometry sub--sample of the catalogue with those derived from other dwarf samples.  The most likely explanation for our systematically higher metallicities compared to all the other samples shown in Figure \ref{fig:mass_metal} is our exclusion of low--metallicity galaxies as a result of our emission line SNR criterion. Applying this criterion  results in the exclusion of an increasing fraction of galaxies as stellar masses decreased and is likely the cause of the apparent flattening of mean metallicity seen in our mass--metallicity relation, below log(M$_*$/\msolar) $\sim$ 8.5.

\item 
 The shallower mass--metallicity relation for our
catalogue is similar to the one reported for recently formed
high–metallicity TDGs \citep{Recchi2015}. This could suggest, at least a fraction of our catalogue dwarfs could have tidal origin. However, we refrain from making any strong claim about this as this is merely a speculation based on just the similarity of slopes from \cite{boquien10,duc14}. Future planned high resolution studies of these dwarfs will reveal more about their formation scenarios.

\item
Our analysis of the impact of  near neighbours reveals the majority of the clean photometry dwarfs  are relatively isolated. However, the few clean photometry dwarfs, with tidally induced SF parameters above the $P_{gg}$ threshold, display bluer g -- i colours than these rest of the clean photometry dwarfs. 

\item 
 The  mean EW(H$\beta$) of the catalogued dwarfs is nearly an order of magnitude lower than for samples of dwarfs with high SF rates, e.g., HII/BCD galaxies.  Moreover, the mean SFR of the catalogued dwarfs is between the high star forming BCDs and low starforming LSBs. This together with  the narrow colour range, and relative isolation of the vast majority of the catalogue's clean photometry dwarfs is consistent with a scenario in which the observed SF in the vast majority of the catalogued dwarfs arises from internally driven episodic mild bursts of SF rather than being driven by external interactions. Although, resolved multi--wavelength observations would be required to confirm this. 
\end{enumerate}

This catalogue will be useful for the following categories of future investigations:   (1) resolved studies of  subsets of the catalogue focused on  metallicity evolution, gas infall/ accretion and SF (2) statistical studies of global properties of nearby dwarfs, especially high metallicity dwarf studies (3) selection of sub--samples for multi wavelength studies (e.g. optical and \hi/CO/ radio continuum).

\label{sec:conclude}

\section*{Acknowledgements}
 We would like to thank the anonymous referee for his/her comments which have led to considerable improvements in the paper.
YG acknowledges support from the National Key Research and Development Program of China (2022SKA0130100), and the National Natural Science Foundation of China (grant No. 12041306). PL (10.54499/DL57/2016/CP1364/CT0010) and TS (10.54499/DL57/2016/CP1364/CT0009) are supported by national funds through Funda\c{c}\~{a}o para a Ci\^{e}ncia e a Tecnologia (FCT) and the Centro de Astrof\'isica da Universidade do Porto (CAUP). This research has made use of the Sloan Digital Sky Survey (SDSS) http://www.sdss.org/. This research has made use of the NASA/IPAC Extragalactic Database (NED) which is operated by the Jet Propulsion Laboratory, California Institute of Technology, under contract with the National Aeronautics and Space Administration.

\section*{Data Availability}


We  used archival SDSS DR--16 data in the catalogue presented in this paper. The full catalogue is published as an online table.



\bibliographystyle{mnras}
\bibliography{cig}




\appendix

\section{A comparison with the MPA--JHU}
\label{comparison}

\begin{figure}
\begin{center}
\includegraphics[scale=.22]{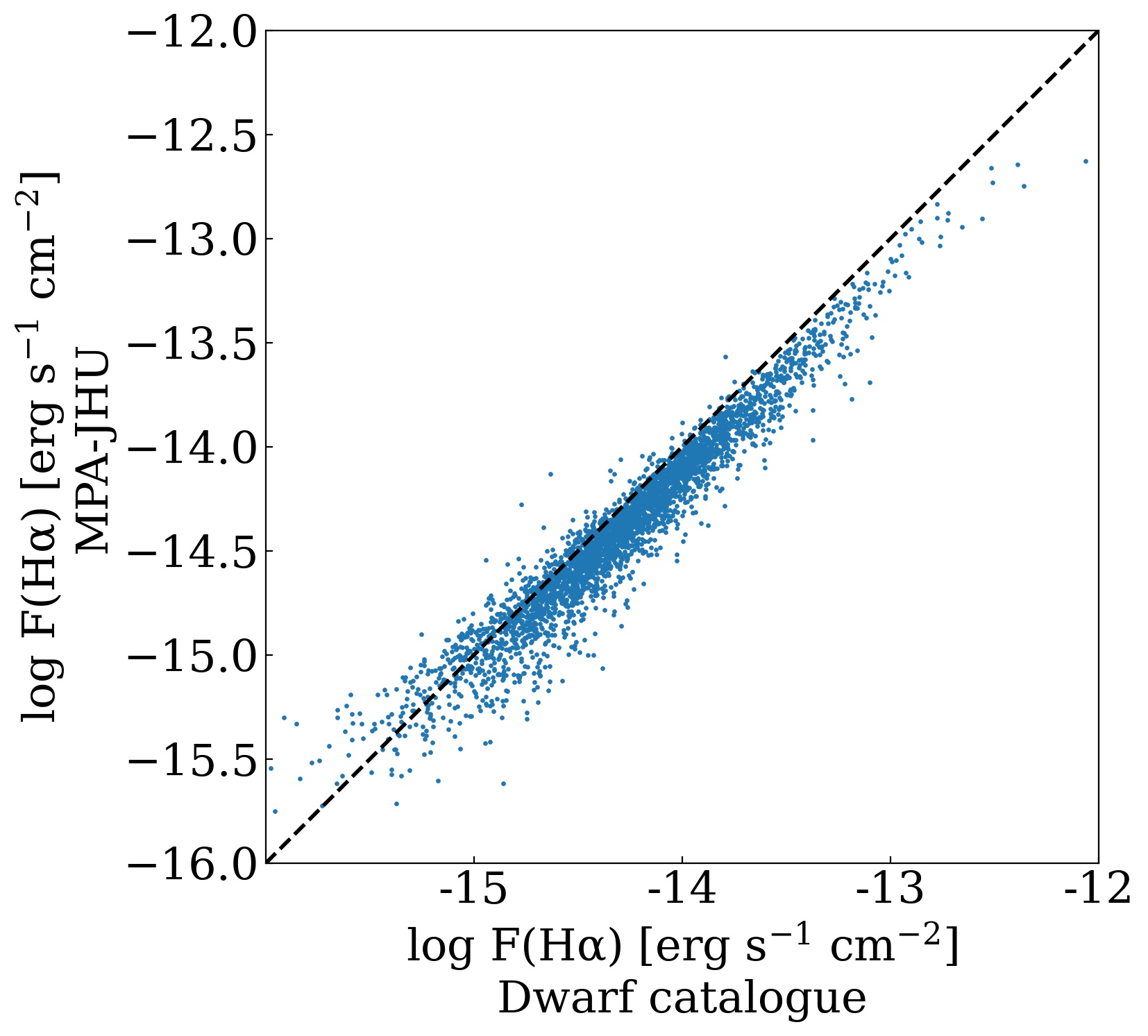}
\includegraphics[scale=.22]{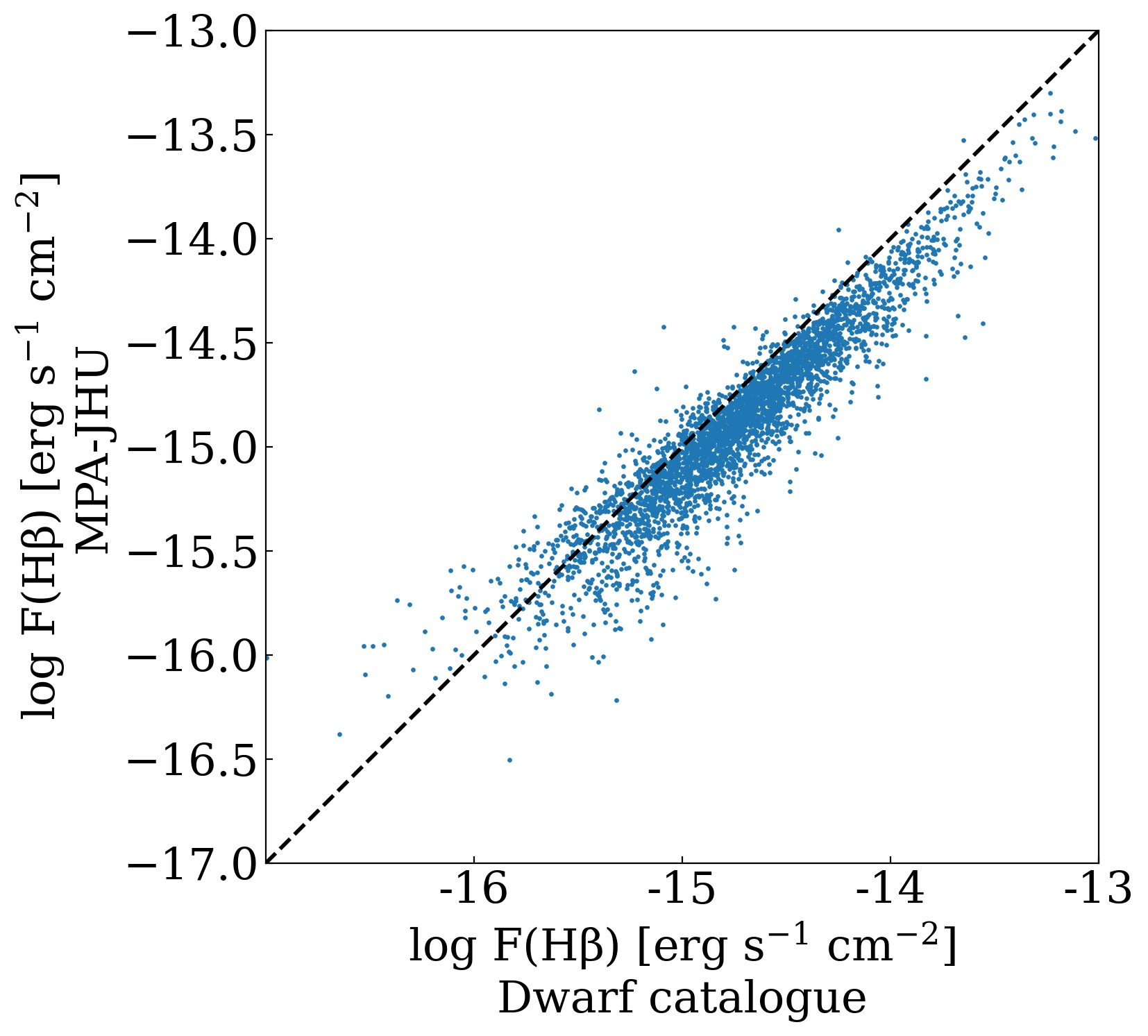}
\includegraphics[scale=.22]{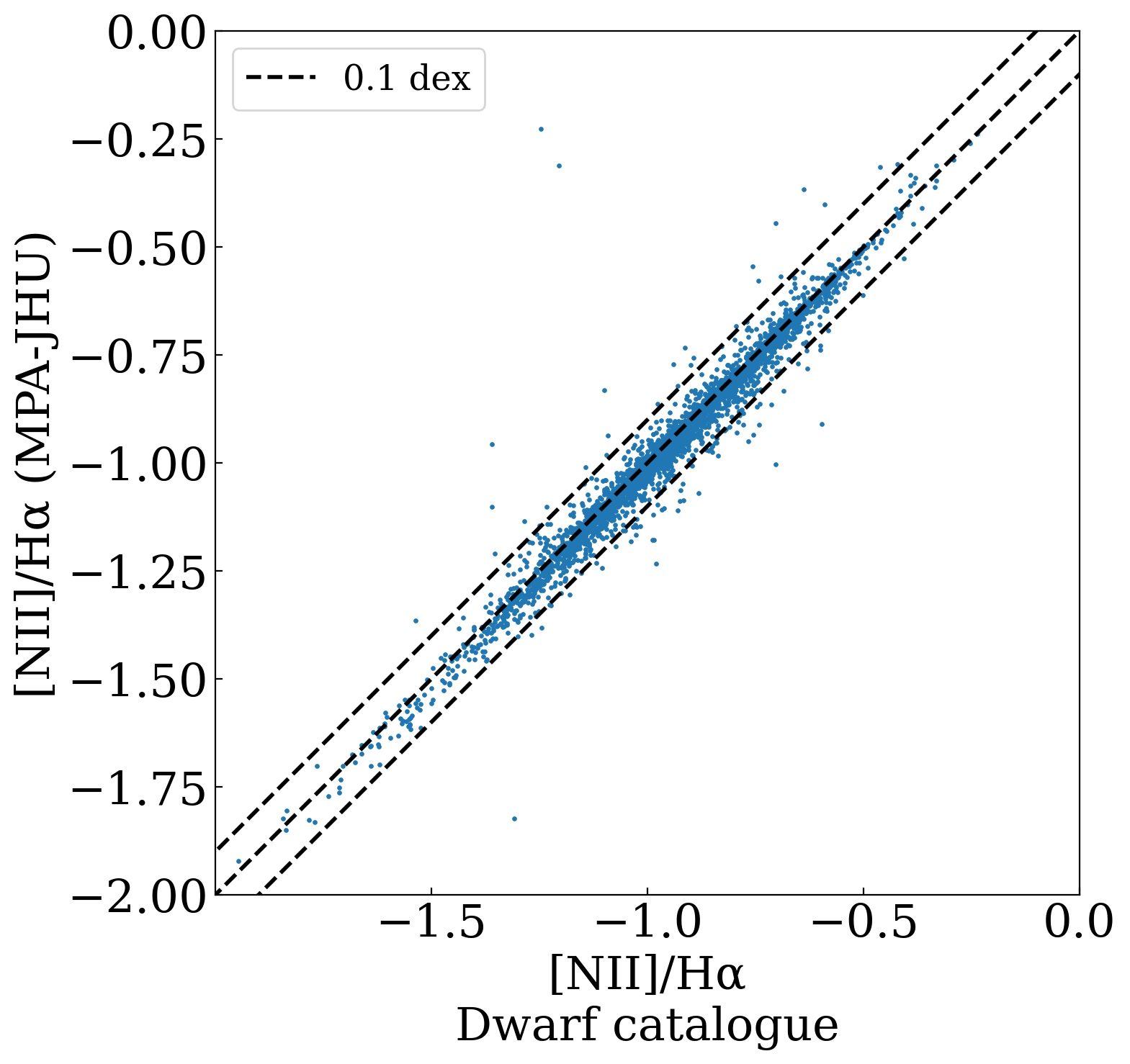}
\includegraphics[scale=.22]{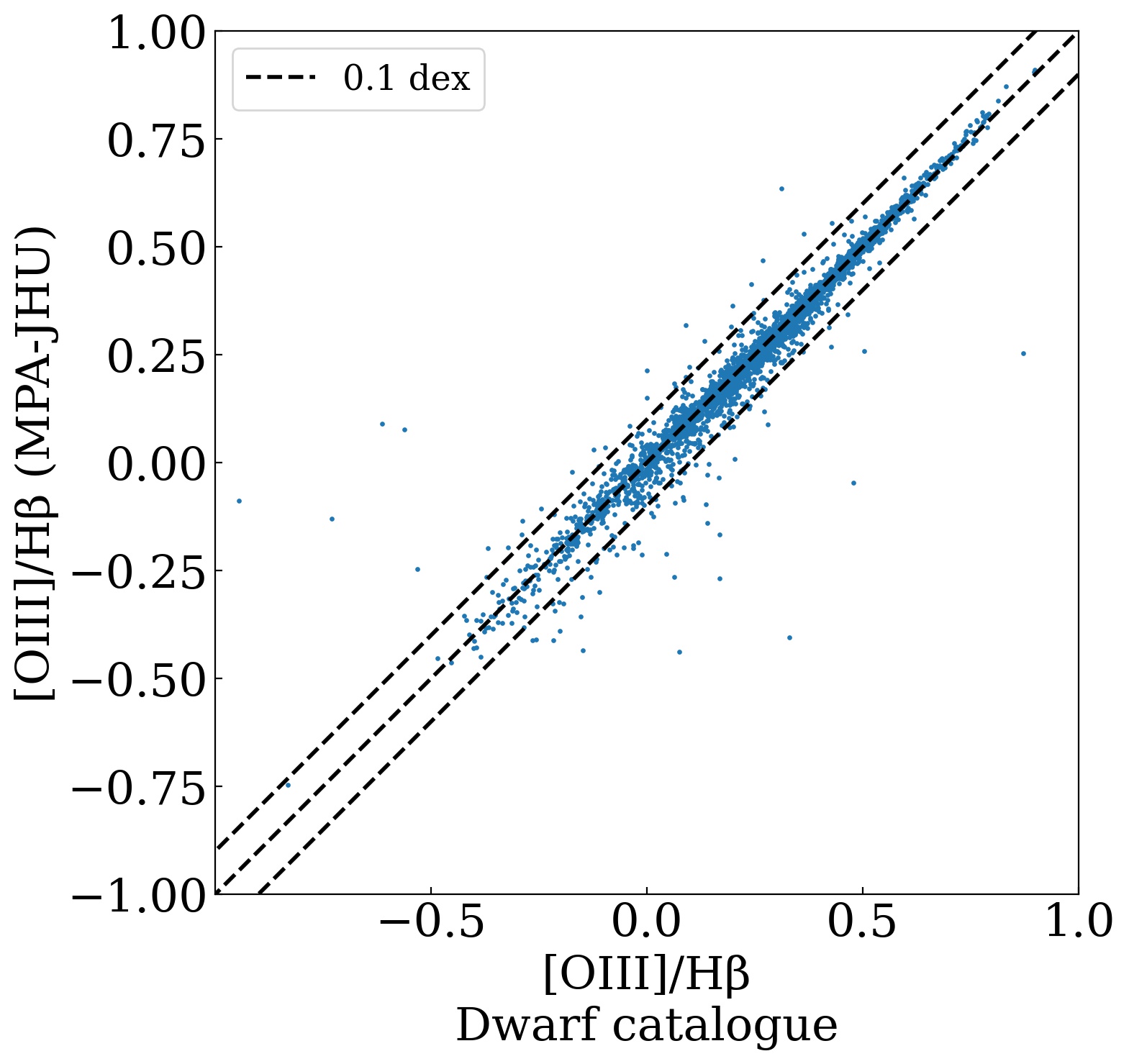}
\end{center}
\caption{\textcolor{black}{Comparison of H$\alpha$ and H$\beta$ fluxes for dwarf galaxies (left) and [NII]/H$\alpha$ and [OIII]/H$\beta$ (right) The x--axes show values derived from the methods described in Section 2  and the  same properties from the MPA--JHU online archive are shown on the y--axis.}}
\label{fig:comparison_flux_ratios_fado_mpa}
\end{figure}

\begin{figure}
\begin{center}
\includegraphics[scale=.27]{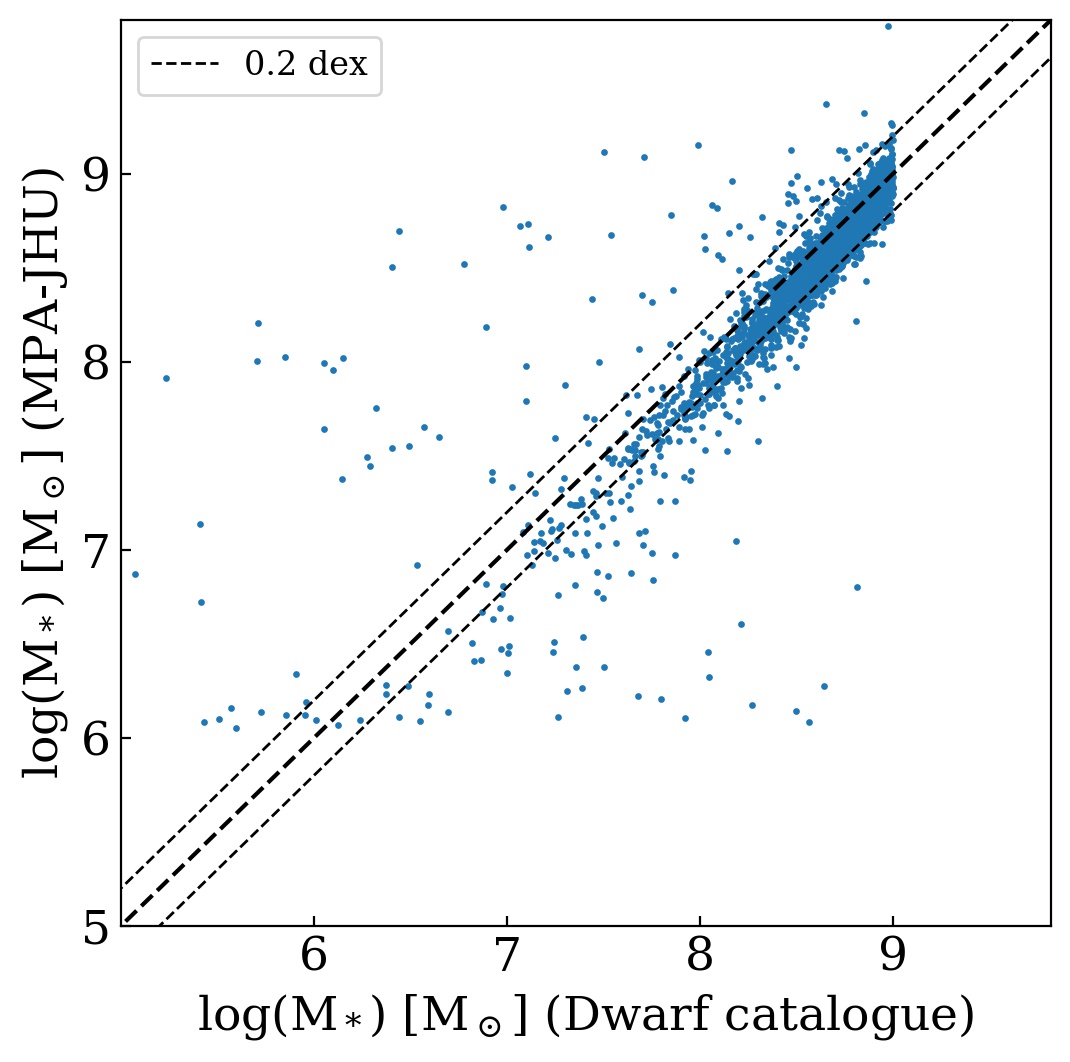}
\includegraphics[scale=.27]{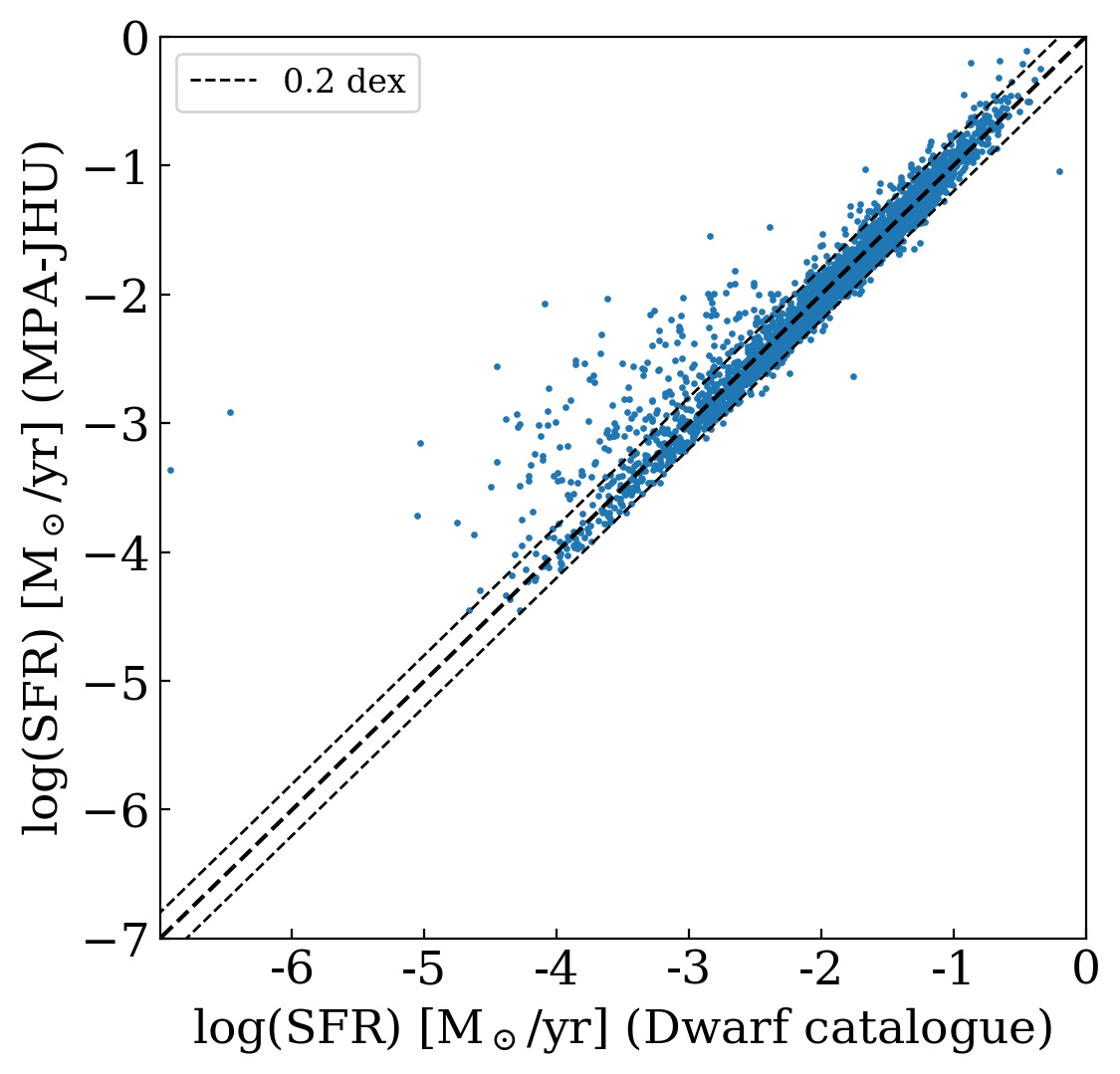}
\includegraphics[scale=.27]{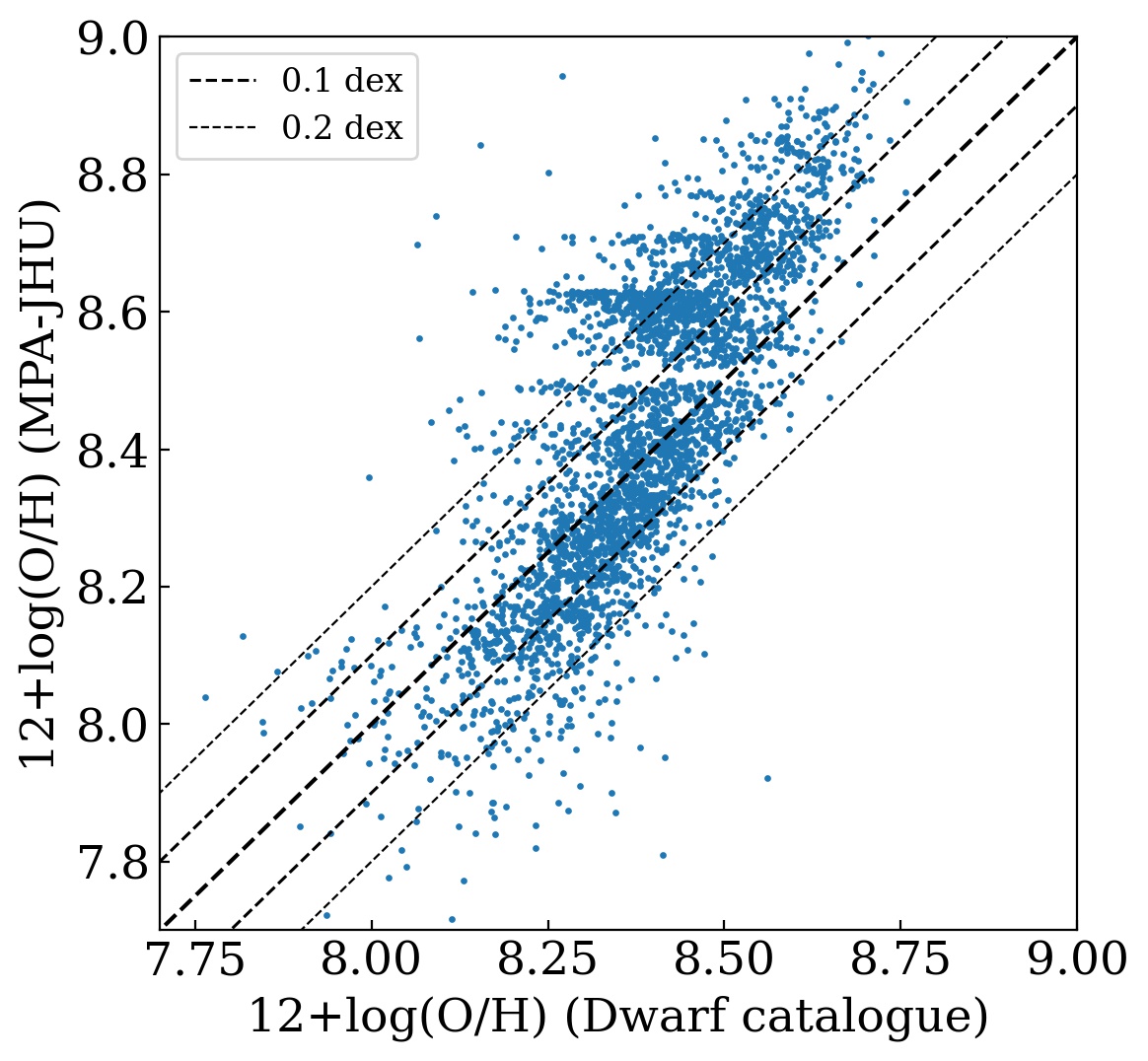}
\caption{\textcolor{black}{Comparison of properties of the dwarf galaxy catalogue derived from the methods described 
    in Section \ref{methods}, \ref{result:sfr} and \ref{result:metal}, respectively. 
    (x -- axis) and the  same properties from the MPA--JHU online archive (y -- axis). 
    From left to right the plots show log(stellar mass), log(SFR(\halpha)) and 12 + log(O/H) oxygen abundance. 
    The central bold line is the 1:1 relation and the lighter dashed lines show the 1 and $\pm$0.1/0.2 dex uncertainties.}}
    \label{fig:comparison}
    \end{center}
\end{figure}

Figure \ref{fig:comparison} compares stellar mass, SFR(H$\alpha$) and 12 + log(O/H) from our dwarf catalogue galaxies in  the MPA--JHU dataset with those derived by the methods described in Section \ref{methods}, \ref{result:sfr} and \ref{result:metal},
respectively. We note that the 12 + log(O/H) in our case corresponds to the average of metallicities 
from the \textcolor{black}{various calibrators described} in Section \ref{result:metal}.

While our sample and the MPA--JHU sample are not exactly comparable because of the different SDSS data releases and Population Spectral Synthesis methods applied, we note that the stellar masses and SFRs have similar trends  and a fraction of the metallicities show differences above 0.2 dex. MPA--JHU results have been estimated from DR7 and DR8 versions of SDSS, both being basically similar \url{https://wwwmpa.mpa-garching.mpg.de/SDSS/} and \url{https://www.sdss4.org/dr12/spectro/galaxy_mpajhu/}. To understand the discrepancy between the MPA--JHU results and our catalogue results, we ran a series of tests to confirm whether this difference originates from using different versions of SDSS or  the different ways  12 + log(O/H) is estimated.

\begin{figure}
\begin{center}
 \includegraphics[scale=.30]{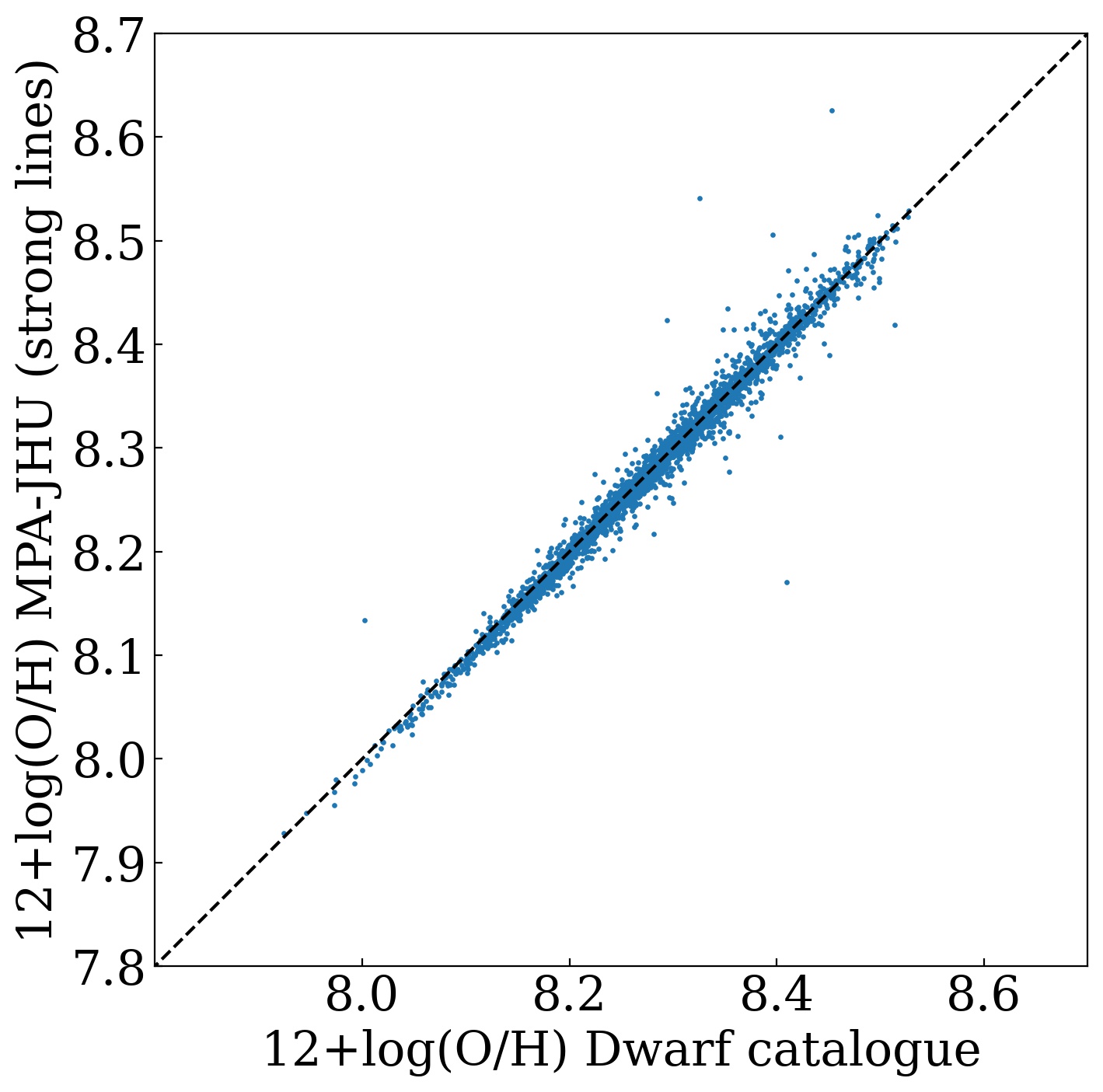}
    \caption{\textcolor{black}{Comparison between 12 + log(O/H) from the dwarf catalogue and from MPA--JHU line fluxes.}}
   \label{fig:met_comparison}
   \end{center}
\end{figure}

To check whether the source data for the MPA--JHU catalogue, has any
systematic difference compared to DR--16, we extracted \textcolor{black}{and measured line fluxes  for the galaxies in our catalogue from DR--8 and DR--16.} In both cases we used FADO to extract and measure the \halpha\ line fluxes. 

As expected, the comparison showed no intrinsic difference between the two data release versions.
\textcolor{black}{Our results are in agreement with the ones found by \cite{Miranda2023}, where they found that 
for star-forming galaxies from the SDSS--DR7, the additional modelling of the nebular contribution, using FADO, does not affect the derived fluxes and consequentially the SFR(H$\alpha$) and stellar masses.}
 
As seen in Figure \ref{fig:comparison}, for \textcolor{black}{the dwarf catalogue} galaxies, our metallicity estimates differ from the MPA--JHU galaxies' estimates in a significant fraction of galaxies. \textcolor{black}{To understand this
difference we carried a series of tests.  We used the flux values from MPA--JHU and applied the calibrator methods applied to our dwarf catalogue. Figure \ref{fig:met_comparison} shows the 12 + log(O/H) derived from both data sets are in good agreement. 
The important point to note is that both methods of estimating metallicities use} line ratios and not absolute fluxes. 
This \textcolor{black}{implies that the difference in metallicities seen in Figure \ref{fig:comparison} likely stems from applying different metallicity calculation methods, rather than a difference in line ratios within the data sets}. Figure \ref{fig:comparison} also shows the calculated stellar masses of the two samples to be consistent. Stellar masses in our sample have been estimated from SDSS DR--16 photometric data and is thus not influenced by line fluxes or the population spectral synthesis code used. 

\section{ONLINE MATERIAL}
\label{online}

\textcolor{black}{The full dwarf galaxy catalogue is available online.}


\bsp	
\label{lastpage}
\end{document}